\documentclass[onecolumn,nofootinbib,amsmath,amssymb,aps]{revtex4-1}
\usepackage{mathtools,cancel}
\usepackage{amsmath}
\usepackage{amsthm}
\usepackage{slashed}
\usepackage{alltt}
\usepackage{soul}
\usepackage{graphicx}
\usepackage{dcolumn}
\usepackage{bm}
\usepackage{color}
\usepackage{mathrsfs}
\usepackage{hyperref}
\usepackage{soul}
\usepackage[dvipsnames]{xcolor}
\usepackage[all]{xy}
\usepackage[titletoc]{appendix}

\hypersetup{
breaklinks=true}

\begin{document}


\title{Asymptotic conformal symmetry at spatial infinity}

\author{Sepideh Bakhoda${}^{1,2}$\footnote{s\_bakhoda@sbu.ac.ir}, 
Fatemeh Mahdieh${}^1$\footnote{f\_mahdieh@sbu.ac.ir}, 
Hossein Shojaie${}^1$\footnote{h-shojaie@sbu.ac.ir}}
\affiliation{${}^1$ Department of Physics, Shahid Beheshti University, G.C., Evin, Tehran 1983969411, Iran\\
${}^2$ Institute for Quantum Gravity, FAU Erlangen-N{\"u}rnburg, Staudstra{\ss}e 7, 
91058 Erlangen, Germany}

\begin{abstract}
   In this paper, the effects of adding spatial conformal symmetry to the asymptotic symmetry group of an asymptotically conformally flat spacetime are studied. It is shown that, in addition to the BMS group, only the dilations of the spatial conformal generators keep the corresponding boundary conditions conformally invariant under hypersurface deformations. We prove that in order to attain (i) a well-defined symplectic structure and (ii) a finite and (iii) integrable conserved charge, these conditions are satisfied simultaneously when admitting Regge-Teitelboim and twisted Henneaux-Troessaert parity conditions, where the latter also contain supertranslation invariance. The conserved dilation charge contains nonzero terms independent of the field variables, giving a nonvanishing effect on the boundary. The dilation symmetry also modifies the ADM mass, which is another physical effect of the conformal symmetry. 
\end{abstract}

\maketitle
\tableofcontents
\newpage
\section{Introduction}

There has been growing interest in the study of boundaries in general relativity in the form of asymptotic symmetries of spacetimes. 
These symmetries also exist at the boundary of asymptotically flat spacetimes, and have shed light on the gravitational effects of gravitating sources. 
Particularly, it was unexpectedly discovered by Bondi, van der Burg, and Metzner~\cite{Bondi:1962px} (and shortly after by Sachs~\cite{Sachs:1962wk,Sachs:1962zza}) that the asymptotic region of an asymptotically Lorentzian flat spacetime, has a far richer structure than the Poincar\'e symmetry group. This symmetry group, dubbed the BMS group, is the trivial extension of the traditional finite-dimensional Poincar\'e group by the infinite-dimensional supertranslations~\cite{Sachs:1962wk}. Strictly speaking, the asymptotic region of an asymptotically flat spacetime serves as an infinite number of classical vacua in GR. In this context, supertranslations characterize the transformations relating these vacua.

There are two different approaches to find the generators of these symmetries. The covariant and the canonical approaches. While the covariant phase space approach has been used for finding the generators of asymptotic symmetries at null and spatial infinity~\cite{Sachs:1962wk,Troessaert:2017jcm}, the latter is preferred to find these generators at spatial infinity. For instance in~\cite{Henneaux-ADMBMS,Henneaux:2018hdj,Henneaux:2019yax}, these generators have been studied and the asymptotic charges have also been calculated in the canonical formalism. Previous studies in this area can be found in ~\cite{Regge:1974zd,Beig1982,Beig1984,Beig:1987zz}.  

In the full four-dimensional case, the true degrees of freedom are similarly constructed by modding out the trivial degrees of freedom, i.e. the trivial diffeomorphisms, from the group of transformations. In this context, large gauge transformations are defined by the extra gauge transformations with nonzero charge, which remain after eliminating the trivial gauge degrees of freedom. A study of the definition of the conserved charges in the covariant formalism can be found, for instance, in ~\cite{BarnichBrandt,WaldZoupas}. These extra symmetries for asymptotically flat spacetimes have been studied by many authors, including their relation to soft theorems and gravitational memory~\cite{Penrose:1962ij,Newman:1962cia,Penrose:1965am,Campiglia:2014yka,Campiglia:2015yka,Campiglia:2016efb,Campiglia:2016jdj,Ashtekar:1996cd,Ashtekar:2014zsa,Andy2014Soft,He2015,Sasha2014,He2014Newsym,Pasterski2016,Kapec:2015vwa}, and well reviewed in~\cite{Madler:2016xju,Alessio:2017lps,Andy-book,Strominger:2017zoo}. It is shown that these spacetimes have an infinite-dimensional symmetry group $BMS = Poincar\Acute{e}\times ST $, where $ST$ stands for supertranslations, which are time translations that represent vacuum transitions between different states of a radiating black hole~\cite{Hawking:2016msc,Compere:2016hzt,Compere:2016gwf,Hawking:2016sgy,Bousso:2017dny,BMSinvGravScat,Strominger:2017aeh,Bousso:2017rsx,Haco:2019ggi,Haco:2018ske}. This symmetry group can be further extended to give an even larger group, which includes superrotations~\cite{Banks:2003vp,Barnich:2009se,Barnich:2010eb,Barnich:2011ct,Barnich:2011mi,Barnich:2013axa,Kapec2014-SR,Flanagan:2015pxa,Barnich:2016lyg,Strominger:2016wns,Barnich:2017ubf,Ball:2019atb}. Superrotations are given by the central extension of the Lie algebra of the symmetry generators of the 2-sphere at infinity. Moreover, Haco \textit{et al.} ~\cite{Haco:2017ekf} have shown that considering conformal symmetry, the spacetime also enjoys asymptotic symmetries at null infinity, namely the conformal BMS group, as they call it. These include the previously found supertranslations and superrotations, as well as dilation, BMS dilations and BMS special conformal transformations. They also compute the conformal BMS algebra. In addition conformal symmetry has been studied in other related topics, for example in~\cite{Pate:2019mfs,Donnay:2018neh}.  

Our interest in including spatial conformal symmetry in the study of the boundary of spacetime is motivated by the fact that it is shown to be included in the natural configuration space of canonical general relativity, known as conformal superspace. The idea of conformal superspace first appeared in \cite{Lichnerowicz}, but it was the work by Barbour and Murchadha \cite{Barbour-ConfSS}, as an extension of the seminal work by York \cite{York:1972sj}, that made it precise. In this context, superspace is the quotient of the space of Riemannian
3-metrics on a 3-manifold by 3-diffeomorphisms and conformal superspace is obtained by modding out three-dimensional conformal transformations from superspace.
As conformal superspace is considered as the natural configuration space of GR, it is conceptually appealing to see the effect of conformal symmetry on its dynamics. It is well known that Einstein gravity is not conformal invariant, in general. Nonetheless, in some cases where the theory has been formulated such that it acquires conformal symmetry~\cite{Oda:2016pok,Oda:2016psn,Jackiw:2014koa}, it has been shown that this symmetry is ``fake'' and does not contribute to the conserved charge. In other words, having a vanishing Noether charge might make one think of the symmetry as a fake symmetry. In another study, it was claimed that these theories, together with Einstein gravity, are locally equivalent and have the same number of symmetry generators, but they differ in their global properties and gauge algebra~\cite{Gielen:2018pvk}. It is noteworthy to mention that boundary terms are not taken into account in these studies.

In this paper, we add spatial conformal symmetries alongside Poincar\'e symmetries, and we use the canonical formalism to implement our analysis. The spacetime considered here is conformally flat at the boundary, that is, spatial infinity; hence the geometry at spatial infinity is asymptotically conformally flat. Considering these new boundary conditions, we find that, in addition to the BMS group with supertranslations found in~\cite{Henneaux-ADMBMS,Henneaux:2018hdj,Henneaux:2019yax}, the boundary conditions are also invariant under hypersurface deformations induced by dilations. This extra asymptotic dilation symmetry gives rise to additional finite and divergent terms in the boundary term. The relevant charges are not integrable, in general; hence we can assume finiteness and integrability to restrict our boundary conditions further. 
Moreover, we find that in contrast to the claims of conformal symmetry being fake in gravity, in some circumstances the boundary terms can indeed give rise to the existence of a nonzero charge related to the spatial conformal symmetry. In particular, the circumstances under which we find this charge are when we consider an asymptotically conformally flat spacetime exhibiting spatial conformal symmetry at the boundary, i.e. spatial infinity. Furthermore, only the dilation generators of the spatial conformal symmetry respect the boundary conditions concurrently.

This paper is organized as follows. In Sec. \ref{Asymptotic symmetries}, a review of asymptotic symmetries at spatial infinity is presented in the framework of the Hamiltonian (ADM) formalism. Specifically, the invariance of the boundary conditions under hypersurface deformations is used to specify the symmetry generators. In Sec. \ref{Adding conformal symmetry}, generators of spatial conformal symmetry are included in the analysis of the boundary of an asymptotically conformally flat spacetime. In addition, hypersurface deformations of the metric are carried out, where we show that the conformal symmetry generators are restricted to dilations only. Section \ref{Asymptotic charges} provides an expression for the conserved charge associated with the symmetry generators, for which the details of the calculation is covered in Appendix \ref{Appendix-Charges}. This section ends with a short discussion on the status of the equations of motion regarding the effects of spatial conformal transformations. In Sec. \ref{Section-Parity}, we briefly review the previously studied Regge-Teitelboim \cite{Regge:1974zd} and Henneaux-Troessaert \cite{Henneaux-ADMBMS,Henneaux:2018hdj,Henneaux:2019yax} parity conditions and examine their application in the case of an asymptotically conformally flat spacetime. In an effort to introduce a new parity condition, we present a theorem proving that, ``if we take $\bar{\pi}^{rr}$ to be \textit{strictly odd}, all attempts to attain (i) a well-defined symplectic structure and (ii) a finite and (iii) integrable conserved charge simultaneously will lead to Regge-Teitelboim (R-T) parity conditions.'' We then show that the original Henneaux-Troessaert (H-T) parity conditions~\cite{Henneaux-ADMBMS} are not suitable for eliminating the divergent term in the boundary term produced by dilation symmetry; however, the new twisted H-T parity conditions~\cite{Henneaux:2018hdj,Henneaux:2019yax} do in fact resolve this issue while maintaining the supertranslations. This is followed by a brief interpretation of the finite terms of the conformal charge in Sec. \ref{Dilation-Charge}. Finally, a discussion is given in Sec. \ref{Conclusion and outlook}.


 \section{Asymptotic symmetries at spatial infinity}\label{Asymptotic symmetries}
 In this section we briefly review the asymptotic symmetries at spatial infinity in an asymptotically flat spacetime. This is done in the context of the Hamiltonian formalism
 of general relativity in which one supposes that spacetime is foliated into a family of spacelike hypersurfaces. In this scenario, the metric tensor of spatial slices, $g_{ij}$, the lapse function $n$, and the shift vector $n^i$, together with their conjugate momenta $\pi^{ij}$, $\Pi$, and $\Pi^i$, respectively, play the roles of the dynamical variables. We define $N = (n,n^i)=(n,\textbf{n})$. After obtaining the ADM action, it turns out that the action does not depend on the time derivatives of $n$ and $n^i$, which leads to the primary constraints $\Pi =0$ and $\Pi^i=0$, respectively, and for the conjugate momentum of the metric, one finds
 \begin{equation}\label{pi}
     \begin{split}
         \pi^{ij}= \sqrt{g}(K^{ij}-{^{(3)}K}g^{ij}),
     \end{split}
 \end{equation}
 where $K_{ij}=\frac{1}{2n}(\dot{g}_{ij}-\mathcal{L}_{\textbf{n}}g_{ij})$ is the second fundamental form of the spatial hypersurface and ${^{(3)}K}$ stands for its trace.
 
 The stability of the primary constraints shows that the secondary constraints are
 \begin{equation}\label{S-C}
     \begin{split}
        &H:=\frac{1}{\sqrt{g}}\left[(g_{ik}g_{jl}-\frac{1}{2}g_{ij}g_{kl})\pi^{ij}\pi^{kl}\right]-\sqrt{g} \;{^{(3)}R} 
        \\
        &H_i:=-2g_{ik}D_j \pi^{jk},
     \end{split}
 \end{equation}
where $g:=\det (g_{ij})$, ${^{(3)}R}$ is the Ricci scalar of the spatial hypersurface, and $D_i$ is the torsion-free, metric compatible connection with respect to $g_{ij}$.
Plugging (\ref{pi}) into (\ref{S-C}), one gets
\begin{equation}\label{C-in terms of K}
    \begin{split}
        &H = \sqrt{g}(K_{ij}K^{ij} - ({^{(3)}K})^2 - \prescript{(3)}{}{R})\\
        &H^i = 2 \sqrt{g} D_j(K^{ij} - {^{(3)}K}g^{ij}).
    \end{split}
\end{equation}
By imposing stability of these constraints under evolution, no tertiary constraints arise. As we will see, the asymptotic behavior of the constraints expressed in (\ref{S-C}) is crucial in the analysis of the conserved charges.
 
Following this brief review, we are ready to jump into the analysis of the asymptotic region using canonical formulation of GR. There are some mathematically rigorous definitions for an asymptotically flat spacetime \cite{DeWitt1, DeWitt2} that are beyond the scope of this paper. Thus, we stick to a simpler definition specifying the asymptotic behavior of the metric as
 \begin{equation}\label{g-ST}
     \begin{split}
        g_{\mu \nu}= \eta_{\mu \nu} +\frac{1}{r}h_{\mu \nu}+O(r^{-2}),
     \end{split}
 \end{equation}
 where $h_{\mu \nu}$ is a tensor on the asymptotic 2-sphere. In order to use the advantage of the Hamiltonian formalism, one needs to know the fall-off behaviors of the variables $g_{ij}$ and $\pi^{ij}$. While there is no fingerprint of the decay behavior of the latter in (\ref{g-ST}), the former follows directly from it. The functionally differentiability of the action and the finiteness of ADM momentum are further requirements that make the asymptotic decay of $\pi^{ij}$ special as
 \begin{equation}\label{BC-Cartesian}
     \begin{split}
         & g_{ij} = \delta_{ij} + \frac{1}{r}\Bar{h}_{ij} + \frac{1}{r^2}h^{(2)}_{ij} + O(r^{-3})\\
         & \pi^{ij} = \frac{1}{r^2} \bar{\pi}^{ij} + \frac{1}{r^3}\pi^{(2)ij} + O(r^{-4}),
     \end{split}
 \end{equation}
 where $\bar{h}_{ij}$, $h^{(2)}_{ij}$, $\bar{\pi}^{ij}$, and $\pi^{(2)ij}$ are again tensor fields on the 2-sphere at spatial infinity. An excellent review of these issues is contained in the book by Thiemann \cite{thiemann_2007}.
 
In their pioneering work \cite{Regge:1974zd}, Regge and Teitelboim have shown that demanding the boundary conditions (\ref{BC-Cartesian}) to stay invariant under hypersurface deformations \cite{Arnowitt:1962hi}, 
\begin{equation}\label{HD}
    \begin{split}
        \delta g_{ij} = & \frac{2n}{\sqrt{g}}(\pi_{ij} - \frac{1}{2}\pi g_{ij}) + \mathcal{L}_{\textbf{n}}g_{ij}\\
        \delta \pi^{ij} = & -n\sqrt{g}(R^{ij} - \frac{1}{2}g^{ij}R) + \frac{1}{2}\frac{n}{\sqrt{g}}(\pi_{mn}\pi^{mn} - \frac{1}{2}\pi^2)g^{ij} \\
        & - \frac{2n}{\sqrt{g}}(\pi^{im}{\pi_m}^j - \frac{1}{2}\pi^{ij}\pi) + \sqrt{g}(D^iD^j n - g^{ij}D_i D^i n )
         + \mathcal{L}_{\textbf{n}}\pi^{ij},
    \end{split}
\end{equation}
 will determine the asymptotic behavior of $n$ and $n^i$ as
 \begin{equation}\label{L&SH-Cartesian}
    \begin{split}
        & n = b_ix^i + a(\textbf{n}) + O(r^{-1})\\
        & n^i = {b^i}_jx^j + a^i(\textbf{n}) + O(r^{-1}),
    \end{split}
\end{equation}
where $b_i$ and $b_{ij}(= -b_{ji})$ are arbitrary constants representing boosts and rotations. In turn, the arbitrary function $a(\textbf{n})$ and arbitrary vector $a^i(\textbf{n})$ represent time and spatial translations, respectively.

Since we wish to study the asymptotic region, it is convenient to work with the spherical coordinates $(r,x^A)$ where $x^A$
describes coordinates on the 2-sphere. In this coordinate, Eq. (\ref{BC-Cartesian}) becomes
\begin{equation}\label{BC-Spherical}
     \begin{split}
         & g_{rr} = 1 + \frac{1}{r}\bar{h}_{rr} + \frac{1}{r^2} h^{(2)}_{rr} + O(r^{-3})\\
         & g_{rA} = \bar{h}_{rA} + \frac{1}{r}h^{(2)}_{rA} + O(r^{-2})\\
         & g_{AB} = r^2 \bar{\gamma}_{AB} + r\bar{h}_{AB} + h^{(2)}_{AB} + O(r^{-1})\\
         & \pi^{rr} = \bar{\pi}^{rr} + \frac{1}{r}\pi^{(2)rr} + O(r^{-2})\\
         & \pi^{rA} = \frac{1}{r}\bar{\pi}^{rA} + \frac{1}{r^2} \pi^{(2)rA} + O(r^{-3})\\
         & \pi^{AB} = \frac{1}{r^2}\bar{\pi}^{AB} + \frac{1}{r^3}\pi^{(2)AB} + O(r^{-4}),
     \end{split}
 \end{equation}
 where $\bar{\gamma}_{AB}$ is the metric on the unit 2-sphere. In attaining the last three equations in (\ref{BC-Spherical}), one should pay attention to the fact that $\pi^{ij}$ is not a tensor field but a tensor density. In addition, without loss of generality, we can assume $\bar{h}_{rA}=0$, which simplifies the calculations done in the next sections considerably. This assumption is valid since $\bar{h}_{rA}=0$ can always be achieved by means of a coordinate transformation \cite{Henneaux-ADMBMS}.
 
In spherical coordinates, Eq. (\ref{L&SH-Cartesian}) reads
\begin{equation}
    \begin{split}
        & n = rb+f + O(r^{-1}), \;\; n^r = W +\frac{1}{r}S + O(r^{-2}), \;\; n^A = Y^A + \frac{1}{r}I^A + 
        O(r^{-2}),
    \end{split}
\end{equation}
where $f$, $W$, and $S$ are arbitrary functions and $I^A$ is a vector field on the asymptotic unit 2-sphere. Here, $b$, an arbitrary function satisfying
\begin{equation}\label{boost}
    \bar{D}_A\bar{D}_B b + b \bar{\gamma}_{AB} = 0,
\end{equation}
is the boost parameter, and $Y^A$ is assumed to be the rotation generator; hence it satisfies
\begin{equation}\label{rotation}
    \mathcal{L}_Y\bar{\gamma}_{AB} = 0.
\end{equation}
Note that the symbol $\bar{D}$ represents the torsion-free connection compatible with $\bar{\gamma}_{AB}$. 

In the following section, we investigate whether one can include conformal generators in (\ref{L&SH-Cartesian}) such that the boundary conditions (\ref{BC-Cartesian}) stay conformally invariant under the hypersurface deformations expressed in (\ref{HD}). 

\section{Adding Conformal Symmetry}\label{Adding conformal symmetry}
In our analysis, we consider hypersurface deformations produced by generators containing conformal symmetry that keep the boundary conditions unchanged. Therefore, we define the lapse and shift in such a way that we include dilation and special conformal transformations, i.e.
\begin{equation}\label{CS-Cartesian}
    \begin{split}
        n =& b_i x^i + a(\textbf{n}) + O(r^{-1})\\
        n^i =& ({b^i}_j+F(\textbf{n})\delta^i_j) x^j \\
        &+ \sum_{k=1,2,3}  \left[V(\textbf{n})(x_j x^j \delta^i_k  - 2x_k x^i) + a^i(\textbf{n})\right] + O(r^{-1}),
    \end{split}
\end{equation} 
where  $F$ and $V$ are arbitrary functions on the unit 2-sphere. Note that in the case where  $F$ and $V$ vanish, we recover the lapse and shift considered by~\cite{Regge:1974zd}, which keep the boundary conditions (\ref{BC-Cartesian}) invariant under hypersurface deformations. Our aim here is to investigate the conditions on these extra terms which keep the boundary conditions conformally invariant under the action of these new generators. 

Accordingly, in spherical coordinates, Eq. (\ref{CS-Cartesian}) can be written as
\begin{equation}\label{CS-Spherical}
    \begin{split}
        & n = rb + f + O(r^{-1}) \\
        & n^r = r^2 Z + r F + W +\frac{1}{r}S +O(r^{-2}) \\
        & n^A = rJ^A + Y^A  + \frac{1}{r}I^A + O(r^{-2}),
    \end{split}
\end{equation}
where $f$, $W$, $F$, and $Z$ are arbitrary functions related to time translation, radial translation, dilation, and special conformal transformations, respectively. Similarly, $J^A$ is an arbitrary vector, also related to special conformal transformations. 
In the following, we explore the conformal invariance of the boundary conditions under hypersurface deformations generated by (\ref{CS-Spherical}).

\subsection{Hypersurface deformations and boundary conditions}\label{HD-BC}
Regarding hypersurface deformations with respect to the lapse and shift defined above, and demanding conformal invariance of the boundary conditions, we find, for the ``$ rA $'' component,
\begin{equation}
    \begin{split}
        \mathcal{L}_{N}g_{rA} = & n\partial_t g_{rA} + n^r \partial_r g_{rA} + n^C\partial_C g_{rA} + g_{ri}\partial_An^i + g_{Ai}\partial_r n^i \\
        = & \frac{2b}{\sqrt{\bar{\gamma}}}\bar{\gamma}_{AC}\bar{\pi}^{rC} - Z h^{(2)}_{rA} + Y^C\partial_C\bar{h}_{rA} + rJ^C\partial_C\bar{h}_{rA} \\
        &+ J^C\partial_Ch^{(2)}_{rA} + F \bar{h}_{rA} + 2 rZ\bar{h}_{rA} + 2Zh^{(2)}_{rA} + r^2 J^C\bar{\gamma}_{AC} \\
        & - I^C\bar{\gamma}_{A} + r J^C\bar{h}_{AC} + J^C h^{(2)}_{AC} + r\partial_AF + \partial_AW + r^2\partial_AZ \\
        & + \bar{h}_{rr}\partial_AF + r \bar{h}_{rr}\partial_AZ + h^{(2)}_{rr}\partial_AZ + \bar{h}_{rC}\partial_AY^C + r \partial_AJ^C\bar{h}_{rC} + h^{(2)}_{rC} \partial_AJ^C + O(r^{-1}) \\
        = & (\alpha_o + O(r^{-1}))(\bar{h}_{rA} + \frac{1}{r}h^{(2)}_{rA} + O(r^{-2})),
    \end{split}
\end{equation}
which gives 
\begin{equation}
    \begin{split}
        & O(r^2) =  r^2(J_A+\partial_AZ)\rightarrow J_A + \partial_AZ = 0\\
        & O(r) =  r(\bar{h}_{CA}J^C+\partial_A F + \bar{h}_{rr}(\partial_A Z)) \rightarrow \bar{h}_{CA}J^C + \partial_A F + \bar{h}_{rr}(\partial_A Z) = 0\\
        & O(1) = \frac{2b}{\sqrt{\bar{\gamma}}}\bar{\gamma}_{AC}\bar{\pi}^{rC} - Z h^{(2)}_{rA} + J^C\partial_C h^{(2)}_{rA}
        + 2Zh^{(2)}_{rA}
        - I^C\bar{\gamma}_{AC} + h^{(2)}_{CA}J^C
        \\
        & \;\;\;\;\;\;\;\;\;\;\;+ \partial_A W + \bar{h}_{rr}(\partial_AF) + h^{(2)}_{rr}(\partial_A Z) + (\partial_A J^C)h^{(2)}_{rC}=0.
    \end{split}
\end{equation}
\\
For the ``$ AB $'' component, taking into account that $n\partial_tg_{AB} = O(r)$, we arrive at
\begin{equation}
    \begin{split}
        \mathcal{L}_{N} g_{AB} = &
         n\partial_t g_{AB} + n^r\partial_r g_{AB} + n^C\partial_C g_{AB} + g_{AC}\partial_B n^C + g_{BC}\partial_A n^C + g_{Ar}\partial_B n^r + g_{Br}\partial_A n^r\\
        = & 2r^2 F\bar{\gamma}_{AB} + 2 r^3 Z\bar{\gamma}_{AB} + r^2 Z \bar{h}_{AB} + r^2 \bar{h}_{rA}\partial_BZ + r^2 \bar{h}_{rB}\partial_AZ \\
        & + r^3(J^C\partial_C\bar{\gamma}_{AB} + \bar{\gamma}_{BC}\partial_AJ^C + \bar{\gamma}_{AC}\partial_BJ^C) + r^2(Y^C\partial_C\bar{\gamma}_{AB}+ \bar{\gamma}_{AC}\partial_BY^C + \bar{\gamma}_{BC}\partial_AY^C)\\
        & + r^2 (J^C\partial_C\bar{h}_{AB} + \bar{h}_{BC}\partial_AJ^C + \bar{h}_{AC}\partial_BJ^C) + O(r)\\
        = & (\alpha_0 + O(r^{-1}))(r^2\bar{\gamma}_{AB} + r\bar{h}_{AB} + O(1)).
    \end{split}
\end{equation}
Hence, comparing orders, we find
\begin{equation}
    \begin{split}
        O(r^3) & = r^3 (  2Z\bar{\gamma}_{AB} + \mathcal{L}_J \bar{\gamma}_{AB}) = 0, \;\;\; \rightarrow \bar{D}_CJ^C = -2Z\\
        O(r^2) & = r^2(2F\bar{\gamma}_{AB} - Z\bar{h}_{AB} - \bar{h}_{rA}\bar{D}_BZ - \bar{h}_{rB}\bar{D}_AZ + \mathcal{L}_Y\bar{\gamma}_{AB} + \mathcal{L}_J\bar{h}_{AB}) = r^2 \alpha_0 \bar{\gamma}_{AB}\\
        \;\;\;\;& \xrightarrow{\bar{h}_{rA} = 0} \;\;\; \mathcal{L}_Y\bar{\gamma}_{AB} + 2 F \bar{\gamma}_{AB} + \mathcal{L}_J\bar{h}_{AB} + Z\bar{h}_{AB} = \alpha_0 \bar{\gamma}_{AB}. \\
    \end{split}
\end{equation}
\\
The ``$ rr $'' component, keeping in mind $n\partial_tg_{rr} = O(r^{-1})$, reads
\begin{equation}
    \begin{split}
        \mathcal{L}_{N}g_{rr} & =
        n\partial_t g_{AB} + n^r\partial_r g_{rr} + n^C\partial_C g_{rr} + 2g_{rC}\partial_r n^C +  2g_{rr}\partial_r n^r \\
        & = -Z\bar{h}_{rr} + J^C\partial_C\bar{h}_{rr} + 2F+4rZ +4Z\bar{h}_{rr} + 2J^C\bar{h}_{rC} + O(r^{-1})\\
        &=\alpha g_{rr} + O(r^{-1})\\
        & \rightarrow O(r)=rZ \; \rightarrow Z=0  \\
        & \rightarrow 2F = \alpha_0 -3 Z\bar{h}_{rr} - J^C\partial_C\bar{h}_{rr} \\
        & \;\;\;\;\;\;\;\;\;\; = \alpha_0 - \mathcal{L}_J\bar{h}_{rr}\\
    \end{split}
\end{equation}
where $\alpha = \alpha_0 + O(r^{-1})$.
As a result, putting all the equations together, one finds
\begin{equation}\label{Z=J=0}
    \begin{split}
        &Z=J^A=\partial_A F=\mathcal{L}_{Y}\bar{\gamma}_{AB}=0\\
        &I_A=\frac{2b}{\sqrt{\bar{\gamma}}}\bar{\gamma}_{AC}\bar{\pi}^{rC}+\partial_A W\\
        &2F=\alpha_0.
    \end{split}
\end{equation}
It is worth noting that having found $Z = J^A = 0$, we can conclude that special conformal transformations cannot keep the boundary conditions conformally invariant under hypersurface deformations. Therefore, the only contribution is due to constant angle-independent dilations.

Using (\ref{Z=J=0}), it is easy to see that (\ref{CS-Spherical}) reduces to
\begin{equation}\label{reduced n}
    \begin{split}
        & n = rb + f + O(r^{-1}) \\
        & n^r = r F + W +\frac{1}{r}S+ O(r^{-2}) \\
        & n^A = Y^A + \frac{1}{r}I^A + O(r^{-2}),
    \end{split}
\end{equation}
where $F$ is an arbitrary constant generating dilations. 

We are now equipped with all the requirements needed to calculate the asymptotic charges which will be carried out in the next section. 

\section{Asymptotic charges}\label{Asymptotic charges}
Before addressing the asymptotic charges, we introduce the asymptotic expansion of the parameters needed in the calculation. To attain a meaningful comparison with the findings of~\cite{Henneaux-ADMBMS}, we proceed to use their notation, in particular,
\begin{equation}
    \begin{split}
        \lambda := & \frac{1}{\sqrt{g^{rr}}} = 1 + \frac{1}{r}\bar{\lambda} + \frac{1}{r^2}\lambda^{(2)} + O(r^{-3})\;;\;\;\; \bar{\lambda}=\frac{1}{2}\bar{h}_{rr},\\
        K^A_B = & -\frac{1}{r} \delta^A_B + \frac{1}{r^2}\bar{k}^A_B + \frac{1}{r^3}k^{(2)A}_B + O(r^{-4})\;;\;\;\; \bar{k}^A_B = \frac{1}{2}\bar{h}^A_B + \bar{\lambda}\delta^A_B + \frac{1}{2}\bar{D}^A\bar{\lambda}_B + \frac{1}{2}\bar{D}_B\bar{\lambda}^A, \;\;\; \bar{k} = \bar{\gamma}_{AB}\bar{k}^{AB}, \\
        \lambda_A := & g_{rA}\; ; \;\;\; \lambda^A = \frac{1}{r^2}\bar{\lambda}^A + \frac{1}{r^3}\lambda^{(2)A} + O(r^{-4}),
    \end{split}
\end{equation}
where $\bar{h}_{rA} = \bar{\lambda}_A$, which can be considered to be zero as mentioned before in Sec. \ref{Asymptotic symmetries}. Also, angular indices of all the barred parameters are raised and lowered by the unit 2-sphere metric $\bar{\gamma}_{AB}$. 

Having known the form of the smearing functions that keeps the boundary conditions intact, we can examine the asymptotic charges by calculating the smeared constraints. The boundary term for a general variation of the smeared constraints is given by
\begin{equation}\label{D3}
    \begin{split}
        \mathcal{B}[\delta g_{ij}, \delta\pi^{ij}] = \int_{\partial\Sigma} d^2 x [-2n^i\delta\pi^r_i &+ n^r \pi^{ij}\delta g_{ij} -2\sqrt{\gamma}n\delta K \\
        & -\sqrt{\gamma}\gamma^{BC}\delta\gamma_{AC}(n K^A_B + \frac{1}{\lambda}(\partial_r n - \lambda^{C}\partial_C n)\delta^A_B)].
    \end{split}
\end{equation}
Correspondingly, the boundary term associated with the generators defined in (\ref{reduced n}) is
\begin{equation}\label{charge}
    \begin{split}
        \mathcal{B} = & r\int_{S^2} d^2x\;\{ -2 F\delta\bar{\pi}^{rr} -2 Y^A\bar{\gamma}_{AB}\delta\bar{\pi}^{rB} -2 \sqrt{\bar{\gamma}} b\delta\bar{k}\}\\
        & + \int_{S^2} d^2x \;\{-2 Y^A \delta[\bar{\pi}^{rB}\bar{h}_{AB} + \bar{\gamma}_{AB}\pi^{(2)rB} + \bar{h}_{rA}\bar{\pi}^{rr}] -2 I^A\bar{\gamma}_{AB}\delta\bar{\pi}^{rB} \\
        & - 2W \delta\bar{\pi}^{rr} - 2 \sqrt{\bar{\gamma}} (f \delta \bar{k} + b \delta k^{(2)})\\
        & + \sqrt{\bar{\gamma}} (f + b \bar{\lambda} + \bar{\lambda}^D\partial_Db)\delta\bar{h} - \sqrt{\bar{\gamma}} b \bar{k}^{AB} \delta\bar{h}_{AB} \\
        & -2 F\delta(\bar{\pi}^{rr}\bar{h}_{rr} + \bar{\pi}^{rA}\bar{h}_{rA}+\pi^{(2)rr})]\\
        & + F( \bar{\pi}^{AB}\delta\bar{h}_{AB} +  \bar{\pi}^{rr}\delta\bar{h}_{rr} + 2\bar{\pi}^{rA}\delta\bar{h}_{rA}) \} + O(r^{-1}).
    \end{split}
\end{equation}
In Appendix \ref{Appendix-Charges}, the calculation can be found in more detail.\footnote{By setting $F=0$ in (\ref{charge}), one expects to recover the boundary term found in \cite{Henneaux-ADMBMS}. However, our expression does not contain the term $-\int_{S^2}d^2x \sqrt{\bar{\gamma}}b\bar{h}\delta\bar{k}$ in this circumstance.} Notice that the dilation symmetry gives one divergent term, namely, $-2rF\int d\Omega \delta\bar{\pi}^{rr}$, in addition to those calculated in~\cite{Henneaux-ADMBMS}. We recall that divergencies also arise in the symplectic structure which makes it ill defined. In addition, the charge corresponding to (\ref{charge}) is not integrable; i.e., $\mathcal{B}$ is not an \textit{exact} one-form in the space of field variables. One possible solution to these issues, as employed before in the literature, is to define parity conditions for the conjugate field variables. We refer to this strategy as the parity method. This will be discussed in detail in Sec. \ref{Section-Parity}.


\subsection{Equations of motion (constraints)}\label{EOM}
In the canonical formalism of general relativity, it is straightforward to check that the Einstein equations are equivalent to the constraints as is obvious from the following relations:
\begin{equation}
    \begin{split}
        G_{\mu\nu}n^{\mu}n^{\nu} = & -\frac{H}{2\sqrt{\det(q)}}\\
        G_{\mu\nu}n^{\mu}q^{\nu}_{\;\;\rho} = & \frac{H_{\rho}}{2\sqrt{\det(q)}},
    \end{split}
\end{equation}
where $n^\mu$ is the unit normal vector to the spatial hypersurface, and $q_{\mu\nu} = g_{\mu\nu} + n_\mu n_\nu$ is the first fundamental form of the hypersurface~\cite{thiemann_2007}. 

The constraints  are used in the process of calculating the conserved charges. Hence, we need to consider the effects of conformal transformations, specifically dilations. Under a spatial conformal transformation $\Tilde{g}_{ij} = \Omega^4 g_{ij}$, the momentum constraint is mapped to itself, and the Hamiltonian constraint in (\ref{C-in terms of K}) transforms to
\begin{equation}\label{Conformal hc}
    \begin{split}
        & 8 {D}_i{D}^i\Omega - {^{(3)}{R}}\; \Omega + \Omega^{-7} {K}^{ij}_{TT}{K}_{ij}^{TT} - \frac{2}{3}\Omega^5 ({^{(3)}K})^2 = 0,
    \end{split}
\end{equation}
where $D$ is the torsion-free metric compatible connection with respect to $g_{ij}$, and ${^{(3)}R}$ is the Ricci scalar constructed from $g^{ij}$. Equation (\ref{Conformal hc}) is known as the Lichnerowicz-York (L-Y) equation. In order for the Hamiltonian constraint to stay conformally invariant, the conformal factor $\Omega$ must be a solution to the L-Y equation. In other words, we must take into account conformal factors that map $(g^{ij},K^{TT}_{ij},{^{(3)}K})$ to a new data set $(\Omega^4 g^{ij}, \Omega^{-2}K^{TT}_{ij}, {^{(3)}K})$ in the conformal superspace configuration space that also satisfies the constraints. Here, $K^{TT}_{ij}$ is the transverse traceless part of $K_{ij}$, namely,
\begin{equation}
    \begin{split}
        K_{ij} = K^{TT}_{ij} + \frac{1}{3}{^{(3)}K}g_{ij}.
    \end{split}
\end{equation}
The solution to (\ref{Conformal hc}) always exists and is unique for a positive $\Omega$ if ${^{(3)}K}\neq 0$, and also its transverse traceless part $K^{TT}_{ij}\neq 0$\footnote{For a detailed discussion see~\cite{Barbour-ConfSS}.}. In this case it is obvious that the constraints are conformally invariant, and we can indeed proceed to use the constraints in~\cite{Henneaux-ADMBMS,Henneaux:2018hdj,Henneaux:2019yax} in our analysis. Since $\delta g_{ij} = \alpha g_{ij}$, we have
\begin{equation}
    \begin{split}
        & \Tilde{g}_{ij} = g_{ij} + \epsilon\delta g_{ij} = (1+\epsilon\alpha)g_{ij} = \Omega^4 g_{ij};\;\; \Omega^4 = e^{\epsilon \alpha} \\
    \end{split}
\end{equation}
where $\epsilon$ is a small constant. Therefore, up to first order in $\epsilon$, Eq. (\ref{Conformal hc}) becomes 
\begin{equation}\label{1st-Order}
    \begin{split}
        &\left(-{^{(3)}{R}} + {K}^{ij}_{TT}{K}_{ij}^{TT} - \frac{2}{3} {^{(3)}K}^2\right)
        + \epsilon\left(2{D}_i{D}^i\alpha - \frac{1}{4}{^{(3)}{R}}\alpha - \frac{7}{4}\alpha {K}^{ij}_{TT}{K}_{ij}^{TT} - \frac{5}{6}\alpha {^{(3)}K}^2\right) = 0
    \end{split}
\end{equation}
The expression in the first parentheses is zero, according to (\ref{C-in terms of K}). In the second parentheses, substituting $K^{ij}_{TT}K^{TT}_{ij}$ in terms of ${^{(3)}R}$ and $({^{(3)}K})^2$ from the Hamiltonian constraints, we obtain
\begin{equation}\label{2nd-Term}
    \begin{split}
        &{D}_i{D}^i\alpha - \left({^{(3)}{R}} + ({^{(3)}K})^2\right)\alpha = 0.
    \end{split}
\end{equation}
This equation shows how $\alpha$ is restricted in order to keep the Hamiltonian constraint conformally invariant. One can show that (\ref{2nd-Term}) is consistent with $\partial_A\alpha_0 = 0$ found earlier in (\ref{Z=J=0}). 


\section{Parity conditions}\label{Section-Parity}
The idea of associating parity conditions with the coefficient functions existing in (\ref{BC-Cartesian}), for the first time, was proposed in \cite{Regge:1974zd} in order to eliminate divergent terms appearing in the boundary term. Furthermore, the finiteness of the symplectic structure with respect to the parity conditions has been studied in a more recent work by Henneaux and Troessaert. To have a well-defined symplectic structure, one assigns parities to field variables in such a way that the kinetic term
\begin{equation}\label{Kinetic}
    \int d^3x \; \pi^{ij} \dot{g}_{ij}   
\end{equation}
whose divergent term is
\begin{equation}\label{Leading-Kinetic}
    \begin{split}
        \int \frac{dr}{r}\int d\theta d\phi \left(\bar{\pi}^{rr}\dot{\bar{h}}_{rr}+\bar{\pi}^{AB}\dot{\bar{h}}_{AB} \right),
    \end{split}
\end{equation}
remains finite. Notice the term $\pi^{rA}\dot{\bar{h}}_{rA}$ does not appear in (\ref{Leading-Kinetic}) since $\bar{h}_{rA}=0$. Hence, parity conditions should be defined such that (\ref{Leading-Kinetic}) and the divergent term in (\ref{charge}) vanish. Moreover, the parity conditions must maintain Poincar\'e transformations among the asymptotic symmetries, and also, known solutions of general relativity such as Schwarzschild and Kerr solutions should obey the parity conditions.

In this section we briefly review the previous parity conditions introduced in the literature by Regge and Teitelboim in~\cite{Regge:1974zd}, and Henneaux and Troessaert in~\cite{Henneaux-ADMBMS,Henneaux:2018hdj,Henneaux:2019yax},  which fulfill all the requirements mentioned above. Thereafter, we explore the results of adopting these parity conditions in the case of an asymptotically conformally flat spacetime. Eventually, we present a theorem regarding the appropriate parity conditions needed to acquire a finite expression for the conserved charge including the contribution from dilation symmetry.

\subsection{R-T parity conditions}
These parity conditions were considered by Regge and Teitelboim in 1974  in order to remove the divergent terms from the asymptotic charge and kinetic term (\ref{Kinetic}). They are given by
\begin{equation}
    \begin{split}
        \bar{h}_{rr} \sim \bar{\pi}^{rA} \sim \bar{h}_{AB} & = even, \;\;\;\;\; \bar{\pi}^{rr} \sim \bar{\pi}^{AB} = odd.
    \end{split}
\end{equation}
Keep in mind that the parity conditions should be preserved under the hypersurface deformations. This requirement is satisfied when 
\begin{equation}\label{f-w-RT}
    W-W_P= even, \; \; \; \; f-f_o = odd,
\end{equation}
as reexpressed in \cite{Henneaux-ADMBMS}. Here $W_P= \Sigma_{m=-1}^1 P^m Y_{1m}(x^A)$ is the odd part of $W$, and the even part of $f$ is just a constant denoted by $f_o$. 

In the expression of the boundary term (\ref{charge}), the divergent terms containing $Y^A\bar{\gamma}_{AB}\delta\bar{\pi}^{rB}$ and $\sqrt{\bar{\gamma}}b\delta\bar{k}$ vanish due to these parity conditions, as intended by R-T. However, these parity conditions over-restrict the system, hence leading to the loss of the extra conserved charges associated with supertranslation symmetry. 

Nonetheless, in the case where we have included asymptotic dilations, the only divergent term associated with this symmetry is
\begin{equation}
    \begin{split}
        & -2rF\delta\bar{\pi}^{rr} = -r\alpha_0\delta\bar{\pi}^{rr}.
    \end{split}
\end{equation}
The condition $\bar{\pi}^{rr} \sim odd$ in the R-T parity conditions forces this divergent term to vanish. Furthermore, the terms including $\delta(\bar{\pi}^{rr}\bar{h}_{rr})$, $\delta (\bar{\pi}^{rA}\bar{h}_{rA})$, $ \bar{\pi}^{AB}\delta\bar{h}_{AB}$, $ \bar{\pi}^{rr}\delta\bar{h}_{rr}$, and $\bar{\pi}^{rA}\delta\bar{h}_{rA}$ also vanish as a result of the R-T parity conditions. Consequently, after including conformal symmetry and the R-T parity condition, we are left with an integrable, nonvanishing charge corresponding to dilation, namely,
\begin{equation}
    \begin{split}
        \mathcal{Q}_{dilation} = \int_{S^2} d^2x \; F\pi^{(2)rr}.
    \end{split}
\end{equation} 
This implies that, with R-T parity conditions in asymptotically conformally flat spacetimes at spatial infinity, dilation symmetry contributes to the conserved charge.


\subsection{H-T parity conditions}
\subsubsection{Original H-T parity}
In their 2017 paper~\cite{Henneaux-ADMBMS}, Henneaux and Troessaert observed that two of the divergent terms included in (\ref{charge}) will vanish with the help of the leading order terms of the constraints (\ref{C-in terms of K}), which are
\begin{equation}\label{H-Constraints}
    \begin{split}
        H  = & -\frac{2}{r}\sqrt{\bar{\gamma}}(\bar{D}_A\bar{D}_B\bar{k}^{AB} - \bar{D}_A\bar{D}^A\bar{k}) + O(r^{-2})\\
        H_A = & - 2 \bar{\gamma}_{AB}(\bar{\pi}^{rB} + \bar{D}_C\bar{\pi}^{BC}) \\
        & - \frac{2}{r} \left[\bar{\gamma}_{AB}\bar{D}_C\pi^{(2)BC} - \partial_A\bar{\lambda}\bar{\pi}^{rr} + \bar{D}_B(\bar{h}_{AC}\bar{\pi}^{BC}) - \frac{1}{2}\bar{\pi}^{BC}\bar{D}_A\bar{h}_{CB}\right] + O(r^{-2}).
    \end{split}
\end{equation}
These constraints vanish, and thus the following relations are imposed on the field variables:
\begin{equation}\label{Constraints}
    \begin{split}
        & \bar{D}_A\bar{D}_B \bar{k}^{AB} - \bar{D}_A\bar{D}^A \bar{k} = 0, \;\;\; \bar{\gamma}_{AB}\bar{\pi}^{rB} + \bar{\gamma}_{AB}\bar{D}_C\bar{\pi}^{BC} = 0.\\
    \end{split}
\end{equation}
 These, together with the Lorentz parameter property, i.e. $\bar{D}_A\bar{D}_Bb + b\bar{\gamma}_{AB} = 0$, remove the divergent terms
 \begin{equation}
     \begin{split}
         r\int_{S^2} d^2x \; (-2Y^A\bar{\gamma}_{AB}\delta\bar{\pi}^{rB} - 2\sqrt{\bar{\gamma}}b\delta\bar{k})
     \end{split}
 \end{equation}
 in (\ref{charge}).
 This gives them the freedom to relax the parity conditions and to present different ones from those of R-T, thereby reviving the supertranslation charge. These H-T parity conditions are
\begin{equation}\label{H-T-Parity}
    \begin{split}
        \bar{\lambda} \sim \bar{\pi}^{AB} = even, \;\;\;\;\; \bar{p} \sim \bar{k}_{AB} \sim \bar{\pi}^{rA} = odd, 
    \end{split}
\end{equation}
where $\bar{\lambda}:=\bar{h}_{rr}/2$ and $\bar{p} := 2(\bar{\pi}^{rr}-\bar{\pi}^A_A)$. The nonvanishing supertranslation charge is thus given by 
\begin{equation}\label{STcharge}
    \begin{split}
        \mathcal{Q}_{Supertranslation} = \int_{S^2} d^2x \; ( 4T\sqrt{\bar{\gamma}}\bar{\lambda} + W \bar{p}),
    \end{split}
\end{equation}
where $T:=f+b\bar{\lambda}$ and $W$ are even and odd arbitrary functions on the unit 2-sphere respectively.\footnote{
It is worth noting that, in~\cite{Henneaux-ADMBMS}, the function $T$ is defined to have an extra term, namely $T=f+b\bar{k}+b\bar{\lambda}$. This is due to the fact that the term $-\int_{S^2}d^2x \sqrt{\bar{\gamma}} b\bar{h}\delta\bar{k}$ in the charge calculated by these authors sabotages its integrability. The inclusion of $b\bar{k}$ in $T$ is a remedy to overcome this dilemma. On the other hand, since we have not found this troublesome term in our calculations (Appendix \ref{Appendix-Charges}), we therefore do not need its presence in the definition of $T$. Nonetheless, this minor disagreement in the two calculations does not effect the final result in an essential way.} The parities of $T$ and $W$ guarantee that H-T parity conditions (\ref{H-T-Parity}) stay invariant under hypersurface deformations.

\subsubsection{Twisted H-T parity}
In their recent work~\cite{Henneaux:2018hdj,Henneaux:2019yax}, the authors introduced another set of boundary conditions, called ``twisted boundary conditions,'' proposing that the field variables can have strict parity up to an improper gauge diffeomorphism, namely 
\begin{equation}\label{H-T2019}
    \begin{split}
        & \bar{\pi}^{rr} - \bar{\pi}^A_A = odd \;\;(strictly),\\
        & \bar{h}_{rr} = even,\;\; \bar{\lambda}_A =odd,\\
        & \bar{\pi}^{rr} = (\bar{\pi}^{rr})^{odd} - \sqrt{\bar{\gamma}}
        \bar{D}_C\bar{D}^C\Psi,\;\; \bar{\pi}^{rA} = (\bar{\pi}^{rA})^{even}-\bar{D}^A\Psi,\\
        & \bar{\pi}^{AB} = (\bar{\pi}^{AB})^{odd} + \sqrt{\bar{\gamma}}(\bar{D}^A\bar{D}^B\Psi - \bar{\gamma}^{AB}\bar{D}_C\bar{D}^C\Psi),\\
        & \bar{h}_{AB} = (\bar{h}_{AB})^{even} + 2(\bar{D}_A\bar{D}_B\Phi + \Phi \bar{\gamma}_{AB}),
    \end{split}
\end{equation}
where $\Psi$ and $\Phi$ are even and odd functions, respectively. These boundary conditions are invariant under BMS transformations and give rise to appropriate charges with the same supertranslation charge given in (\ref{STcharge}) and also include, for instance, the electromagnetism charge found in ~\cite{Henneaux:2018hdj}. 
\vspace{0.25cm}

However, when including dilation, the original H-T parity conditions fail to remove the divergent term associated with dilation in the boundary term; $F\delta \bar{\pi}^{rr}$ is not strictly odd with these parity conditions.

On the other hand, considering the twisted H-T parity conditions in the case of an asymptotically conformally flat spacetime, one can see that the form of the even part of $\bar{\pi}^{rr}$, which is $-\sqrt{\bar{\gamma}}\bar{D}_C \bar{D}^C \Psi$, is not invariant under hypersurface deformations. In fact, the even part of $\delta \bar{\pi}^{rr}$, with twisted H-T parity, is 
\begin{equation}
    \begin{split}
        \delta \bar{\pi}^{rr}_{even}
        =&\mathcal{L}_Y \bar{\pi}^{rr}_{even}- F{\pi}^{rr}_{even}+\sqrt{\bar{\gamma}}[- \bar{D}^2f_{even} + (\bar{D}_E\bar{h}^{CE}_{odd})(\partial_Cb) - \frac{1}{2}(\bar{D}^C\bar{h}_{odd})(\partial_Cb) -\frac{1}{2} b\bar{h}_{odd}]\\
        =&\mathcal{L}_Y \bar{\pi}^{rr}_{even}- F{\pi}^{rr}_{even}
        +\sqrt{\bar{\gamma}}[- \bar{D}^2f_{even} + 2\bar{D}_E(\bar{D}_A\bar{D}_B\Phi + \Phi \bar{\gamma}_{AB})(\partial_Cb) - \bar{D}^C(\bar{D}^2\Phi + 2\Phi)(\partial_Cb) - b(\bar{D}^2\Phi + 2\Phi)]
    \end{split}
\end{equation}
On the other hand, considering the twisted H-T parity conditions in the case of an asymptotically conformally flat spacetime, the integration of $2\sqrt{\bar{\gamma}}\bar{D}_C\bar{D}^C\Psi$ over the 2-sphere vanishes. This completely eliminates the divergency of the boundary term, while the divergent terms in the symplectic structure also vanish since the constraints used to eliminate them are conformally invariant, giving a finite dilation charge alongside the supertranslation charges. In the next subsection we will analyze the desired requirements to support the existence of conformal symmetry in a consistent way.

\subsection{Parity conditions appropriate for including dilation symmetry}

There are two scenarios that we examine here in order to find the appropriate parity conditions. First, we take $\bar{\pi}^{rr}$ to be strictly odd and find that this will lead to the R-T parity conditions. In the second case, we examine the twisted H-T parity conditions, where $\bar{\pi}^{rr}$ is odd up to the Laplacian of a function. In what follows, where it is needed, the functions are split into their even and odd parts and denoted by the corresponding subscripts. 

\begin{center}
    \textbf{Taking $\bar{\pi}^{rr}$ to be strictly odd}
\end{center}
\textbf{Theorem:} \; Consider the study of asymptotic symmetries at spatial infinity in the context of the canonical formalism for an asymptotically conformally flat spacetime. If one wishes to use the parity method, it can be shown that by including conformal symmetry in the asymptotic region, only R-T parity conditions can simultaneously satisfy the following three conditions:
\begin{enumerate}
    \item The symplectic structure is well defined. 
    \item The conserved charge is finite. 
    \item The conserved charge is integrable. 
\end{enumerate}

\begin{proof}
Given that the only remaining\footnote{Recall that the other two divergent terms are eliminated by use of the constraints and properties of the Lorentz parameter as mentioned in the previous subsection. } divergent term in the boundary term is expressed by $-2rF\int d^2x\; \delta\bar{\pi}^{rr}$, to satisfy condition 2, one can require $\bar{\pi}^{rr}$ to be a strictly odd density function. To sustain consistency, this parity condition must be invariant under hypersurface deformations, that is 
\begin{equation}\label{delta pi rr}
    \begin{split}
        \delta\bar{\pi}^{rr} = \mathcal{L}_Y \bar{\pi}^{rr}- F\bar{\pi}^{rr} + b\sqrt{\bar{\gamma}} (8\bar{\lambda} + \bar{D}^2\bar{\lambda}) + \sqrt{\bar{\gamma}}[-\bar{D}^2f + 2(\bar{D}_E\bar{k}^{CE})(\partial_Cb) - (\bar{D}^C\bar{k})(\partial_Cb) - b\bar{k}].
    \end{split}
\end{equation}
Accordingly, in order to eliminate the terms that do not respect this parity condition, namely,
\begin{equation}
    \begin{split}
         \sqrt{\bar{\gamma}}[8b\bar{\lambda}_{odd} + b \bar{D}^2\bar{\lambda}_{odd} - \bar{D}^2f_{even} + 2(\bar{D}_E\bar{k}^{CE}_{odd})(\partial_Cb) - (\bar{D}^C\bar{k}_{odd})(\partial_Cb) - b\bar{k}_{odd}],
    \end{split}
\end{equation}

$f_{even}$ should be defined as 

\begin{equation}\label{f_even}
     \begin{split}
         f_{even}(\hat{\textbf{r}}) := & \int_{S^2} d\Omega' \; G(\hat{\textbf{r}},\hat{\textbf{r}}') \left[8b\bar{\lambda}_{odd} + b \bar{D}^2\bar{\lambda}_{odd} + 2(\bar{D}_E\bar{k}^{CE}_{odd})(\partial_Cb) - (\bar{D}^C\bar{k}_{odd})(\partial_Cb) - b\bar{k}_{odd}\right](\hat{\textbf{r}}') + f_o,
     \end{split}
 \end{equation}
where $f_o$ is a constant, $\hat{\textbf{r}}$ and $\hat{\textbf{r}}'$ are unit vectors, and 
\begin{equation}\label{GreenFunction}
     \begin{split}
         G(\hat{\textbf{r}},\hat{\textbf{r}}') = \frac{1}{4\pi} \ln(1 -\hat{\textbf{r}}\cdot\hat{\textbf{r}}')
     \end{split}
 \end{equation}
is the Green function of the Laplace-Beltrami operator $\bar{D}^2$ on the unit 2-sphere.
It is worth noting that by defining 
\begin{equation}
    \begin{split}
        A:=8b\bar{\lambda}_{odd} + b \bar{D}^2\bar{\lambda}_{odd} + 2(\bar{D}_E\bar{k}^{CE}_{odd})(\partial_Cb) - (\bar{D}^C\bar{k}_{odd})(\partial_Cb) - b\bar{k}_{odd},
    \end{split}
\end{equation}
and taking into consideration~(\ref{GreenFunction}), one can rewrite Eq. (\ref{f_even}) as

\begin{equation}\label{f_even_re}
    \begin{split}
         f_{even}(\hat{\textbf{r}})-f_o :=  \int_{S^2}d\Omega' \;& \frac{1}{4\pi}\ln(1 -\hat{\textbf{r}}\cdot\hat{\textbf{r}}')
          A(\hat{\textbf{r}}')
         =\int_{\text{hemisphere}} d\Omega' \; \frac{1}{4\pi} \ln\left[(1 -\hat{\textbf{r}}\cdot\hat{\textbf{r}}')(1 +\hat{\textbf{r}}\cdot\hat{\textbf{r}}')\right]A(\hat{\textbf{r}}')
    \end{split}
 \end{equation}
which shows that $f_{even}(\hat{\textbf{r}})$ is invariant under antipodal mapping on the 2-sphere, consistent with our assumption about its even parity.

In an effort to satisfy condition 1, one can assign the parity conditions of the field variables in such a way so as to enforce the elimination of the divergent part of the kinetic term, i.e. (\ref{Leading-Kinetic}). The two alternative parity conditions available are given in the following cases.  

\textit{Case 1:} If $\bar{h}_{rr}= even$, R-T parity conditions are restored. This is due to the fact that this assumption removes the first divergent term of (\ref{Leading-Kinetic}), and one is left with $\int d\theta d\phi \; \bar{\pi}^{AB}\dot{\bar{h}}_{AB}$. Hereafter, the only option for the remaining parities are $\bar{\pi}^{AB}= odd$ and $\bar{h}_{AB}= even$. Notice that any alternatives will place additional algebraic restrictions on field variables, which is undesirable since the only intended restrictions are parity conditions. Furthermore, one cannot exchange their parity behaviors to $\bar{\pi}^{AB}= even$ and $\bar{h}_{AB}= odd$, seeing that this alternative choice will not preserve the parity of $\bar{h}_{rr}$ under hypersurface deformations, as one can easily observe in Eq. (\ref{HD-of-Lam}). With $\bar{h}_{rr}\sim \bar{h}_{AB}=even$, $\bar{\pi}_{rr}\sim \bar{\pi}_{AB}=odd$, from (\ref{f_even}) we can see $f_{even}=f_o$. Therefore, (\ref{HD-of-pi-rA}) shows that $\bar{\pi}^{rA}$ has to be even.

\textit{Case 2:} If $\bar{h}_{rr}\neq even$, then $\bar{\lambda}_{odd}$ is an arbitrary odd function which is not identically zero, and as will be shown in the following, it is in contradiction with condition 3. In order to proceed, consider the terms which \textit{seem} to jeopardize integrability, i.e.
\begin{equation}
    \begin{split}
        &-2 \left( \frac{2b}{\sqrt{\bar{\gamma}}}\bar{\pi}^{rA} + \bar{D}^AW \right) \bar{\gamma}_{AB}\delta\bar{\pi}^{rB} - 2W\delta\bar{\pi}^{rr} + \sqrt{\bar{\gamma}}\left(-2f\delta\bar{k} + f\delta\bar{h} + b\bar{\lambda}\delta\bar{h} - b\bar{k}^{AB}\delta\bar{h}_{AB}\right) \\
        &\;\;\; + F\left(\bar{\pi}^{rr}\delta\bar{h}_{rr} + \bar{\pi}^{AB}\delta\bar{h}_{AB}\right) .
\end{split}
\end{equation}
Using $\bar{h}_{AB}=2(\bar{k}_{AB}-\bar{\lambda}\bar{\gamma}_{AB})$, we find
\begin{equation}\label{rem}
    \begin{split}
        & \delta\left(-\frac{2b}{\sqrt{\bar{\gamma}}}\bar{\pi}^{rA}\bar{\pi}^r_A\right) - 2 (\bar{D}_AW)\delta\bar{\pi}^{rA} - 2W\delta\bar{\pi}^{rr} + \sqrt{\bar{\gamma}} \left[- 4f\delta\bar{\lambda} + 2\delta(b\bar{\lambda}\bar{k}) - 2\delta(b\bar{\lambda}^2) - \delta(b\bar{k}^{AB}\bar{k}_{AB})\right]\\ 
        &\; \; \;+ F\left(2\bar{\pi}^{rr}\delta\bar{\lambda}+ \bar{\pi}^{AB}\delta\bar{h}_{AB}\right),\\
    \end{split}
\end{equation}
where we have taken  $\bar{\lambda}^A = 0$. Hence, the terms that are inconsistent with condition 3 are 
\begin{equation}\label{remanent}
    \begin{split}
        & \int d^2x \; \{- 2(\bar{D}_A W_{odd})\delta\bar{\pi}_{even}^{rA} - 2(\bar{D}_A W_{even})\delta\bar{\pi}_{odd}^{rA} - 2 W_{odd}\delta\bar{\pi}^{rr}
        +2(F\bar{\pi}^{rr}- 2\sqrt{\bar{\gamma}} f_{odd})\delta\bar{\lambda}_{odd}\\
        &\; \; \; \; \;\;\;\; \; \; \;
        -4\sqrt{\bar{\gamma}} f_{even}\delta\bar{\lambda}_{even}+F\bar{\pi}^{AB}_{odd}\delta\bar{h}^{odd}_{AB}+F\bar{\pi}^{AB}_{even}\delta\bar{h}^{even}_{AB}\}.
    \end{split}
\end{equation}
Now, it is obvious that the term
\begin{equation}\label{lambda}
    -4\int d^2x \; (\sqrt{\bar{\gamma}} f_{even})\delta\bar{\lambda}_{even}=
    -4\int d^2x \; \delta \left(\sqrt{\bar{\gamma}} f_{even}\bar{\lambda}_{even}\right) +4 \int d^2x \; (\sqrt{\bar{\gamma}} \bar{\lambda}_{even})\delta f_{even}
\end{equation}
in~(\ref{remanent}) cannot be canceled out by other terms;  thus the only remaining option is to consider $\int d^2x \; (\sqrt{\bar{\gamma}}\bar{\lambda}_{even}) \delta f_{even}=0$. 

Recall that $f_{even}$ expressed by (\ref{f_even}) is a functional of $\bar\lambda_{odd}(=\frac{1}{2}\bar{h}^{odd}_{rr})$, and therefore it is an arbitrary even function due to the assumption on $\bar{\lambda}_{odd}$. By this means, $\bar{\lambda}_{even}$ has to vanish identically. However, this yields inconsistency with condition 1, and in fact Eq. (\ref{Leading-Kinetic}) will only vanish when the even part of $\bar{\pi}^{AB}\dot{\bar{h}}_{AB}$ cancels $\bar{\pi}^{rr}\dot{\bar{h}}_{rr}$.
On the other hand, this is not desirable as it will place an extra constraint on the field variables. Consequently, the assumption on the parity of $\bar{h}_{rr}$($\neq even$) is inconsistent and $\bar{h}_{rr}$ must be even. This brings us back to case 1, in which recovering R-T parity conditions is inevitable. 
\end{proof}

\begin{center}
    \textbf{Considering the twisted H-T parity conditions regarding $\bar{\pi}^{rr}$}
\end{center}
Although we proved above that the supertranslations vanish when trying to parallelly satisfy the three conditions in the theorem, there is yet another side to this story. Particularly, taking $\bar{\pi}^{rr}$ to be strictly odd is in fact overconstraining the boundary conditions and therefore leads to the loss of supertranslation invariance when dilation symmetry is included. On the other hand, if one considers the more desirable twisted H-T boundary conditions, where $\bar{\pi}^{rr}$ is odd up to the Laplacian of a function, as is obvious in (\ref{H-T2019}), the divergent term associated with dilation symmetry will indeed vanish since its integration over the 2-sphere is zero.\\
Correspondingly, using (\ref{H-T2019}), the transformation of $\bar{\pi}_{even}^{rr}$ under hypersurface deformations is
\begin{equation}
    \begin{split}
         \delta \bar{\pi}^{rr}_{even} = & \sqrt{\bar{\gamma}}\mathcal{L}_Y (\bar{D}^2 \Psi) - F\sqrt{\bar{\gamma}}\bar{D}^2 \Psi + \sqrt{\bar{\gamma}}[- \bar{D}^2f_{even} + (\bar{D}_E\bar{h}^{CE}_{odd})(\partial_Cb) - \frac{1}{2}(\bar{D}^C\bar{h}_{odd})(\partial_Cb) - \frac{1}{2}b\bar{h}_{odd}]\\
         := & \sqrt{\bar{\gamma}}\mathcal{L}_Y (\bar{D}^2 \Psi) + \sqrt{\bar{\gamma}}[-F\bar{D}^2 \Psi - \bar{D}^2f_{even} + \Upsilon ] . 
    \end{split}
\end{equation}
Seeing that $Y^A$ is the Killing vector field on the unit 2-sphere, the term
\begin{equation}
    \begin{split}
        \int_{S^2} d^2x \mathcal{L}_Y(\bar{D}^2\Psi) = & \int_{S^2} d^2x  Y^A\bar{D}_A\bar{D}^2\Psi = \int_{S^2} d^2x \bar{D}_A(Y^A\bar{D}^2\Psi) = 0
    \end{split}
\end{equation}
does not contribute in the divergency. Notice that $\Upsilon$ can also be expressed as $\bar{D}^2\Theta$ in reference to~\cite{Henneaux:2019yax}. Accordingly, all the remaining terms, namely, $-F\bar{D}^2\Psi - \bar{D}^2f_{even} + \bar{D}^2\Theta$, are Laplacians of even functions as desired. In this case, the divergency is eliminated, and we conclude that using the twisted H-T parity, in addition to the supertranslation charge, a contribution from dilation symmetry is also present. 

\section{Evaluation of the dilation charge}\label{Dilation-Charge}
As stated in the proof of the theorem, with R-T parity, the charge is integrable, i.e.
\begin{equation}
    \begin{split}
        \mathcal{B}= -\delta \mathcal{Q}
    \end{split}
\end{equation}
where
\begin{equation}
    \begin{split}
        \mathcal{Q}=& \int_{S^2} d^2x \;[2 Y^A (\bar{\pi}^{rB}\bar{h}_{AB} + \bar{\gamma}_{AB}\pi^{(2)rB})
        + 2W_P (\bar{\pi}^{rr}- \bar{D}_A\bar{\pi}^{rA}) + 2b \sqrt{\bar{\gamma}}k^{(2)}
        +4 f_o\sqrt{\bar{\gamma}} \bar{\lambda} 
        +2 F\pi^{(2)rr}]
    \end{split}
\end{equation}
is the conserved charge at spatial infinity. The term containing $ 2F\pi^{(2)rr} $ is related to the conserved charge resulting from dilation symmetry. To understand this term we can write $\pi^{(2)rr}$($=O(r^{-1})|_{\pi^{rr}}$) using (\ref{pi}), as\footnote{The calculation can be found in Appendix \ref{pi^rr}.}

\begin{equation}
    \begin{split}
        \pi^{(2)rr} = & -\frac{\sqrt{\bar{\gamma}}}{2b^2}\left[ b(\bar{\gamma}^{AB}\dot{h}^{(2)}_{AB} - 4S - W\bar{h} - {\gamma}^{AB}\mathcal{L}_Y  h^{(2)}_{AB} - {\gamma}^{AB}\mathcal{L}_I\bar{h}_{AB}) \right.\\
        & - (2f\bar{\gamma}^{AB} - 2b\bar{k}^{AB})( \dot{\bar{h}}_{AB} + 2 W \bar{\gamma}_{AB} + F\bar{h}_{AB}+ \mathcal{L}_Y\bar{h}_{AB} + \mathcal{L}_I\bar{\gamma}_{AB}) \\
        & \left.+ 2F(b\bar{\gamma}_{AB}h^{(2)AB} + 2b\bar{\lambda}\bar{h} - 6b\bar{\lambda}^2 + 4f\bar{k})\right],
    \end{split}
\end{equation}
which gives
\begin{equation}\label{dilationcharge}
    \begin{split}
        \mathcal{Q}_{dilation}= & \int_{S^2} d^2x \; \sqrt{\bar{\gamma}}\frac{F}{b^2 }\left[ b(\bar{\gamma}^{AB}\dot{h}^{(2)}_{AB} - 4S - W_P\bar{h} - \bar{\gamma}^{AB}\mathcal{L}_Y  h^{(2)}_{AB} - \bar{\gamma}^{AB}\mathcal{L}_I\bar{h}_{AB}) \right.\\
        & - 2f_o\bar{\gamma}^{AB} \dot{\bar{h}}_{AB} -8 f_o W_{even}-8 f_{odd} W_{P} - 2f_oF\bar{h}- 2f_o\mathcal{L}_Y\bar{h} - 2f\bar{\gamma}^{AB}\mathcal{L}_I\bar{\gamma}_{AB} \\
        & \left.+   4b W_P\bar{k}  + 2b\bar{k}^{AB}\mathcal{L}_I\bar{\gamma}_{AB}+ 2F\bar{\gamma}_{AB}(b h^{(2)AB} + 4f_o\bar{k}^{AB})\right]. 
    \end{split}
\end{equation}
The terms containing $\dot{\bar{h}}_{AB}$ and $\dot{h}^{(2)}_{AB}$ are suggestive of traces of radiative effects in the system on the boundary. Although not the same, $\dot{\bar{h}}_{AB}$ resembles the Bondi news tensor\footnote{In the 4-dim scenario, the Bondi news tensor is defined as $N_{AB}=\partial_uC_{B}$, where ``$u$'' is the retarded time coordinate and $C_{AB}$ is the coefficient of the $O(r^{-1})$ term in the asymptotic expansion of the angular metric at null infinity, describing the time dependency of the radiation at null infinity.} in the full 4-dim analysis using the covariant formalism. By use of (\ref{Z=J=0}), it is also evident from this expression that the terms 
\begin{equation}
    \begin{split}
        & \int_{S^2} d^2x \; \sqrt{\bar{\gamma}}\frac{F}{b^2 }\left[ - 4S -8 f_o W_{even}-8 f_{odd} W_{P} - 2f\bar{\gamma}^{AB}\mathcal{L}_{\bar{D}W}\bar{\gamma}_{AB} \right]
    \end{split}
\end{equation}
are independent of the field variables and, in general, nonzero, even in the case of vanishing field variables. One can therefore infer that the existence of a nonzero charge corresponding to the dilation generator is essentially inevitable. This distinguishes our result from the previous understanding of the effect of conformal symmetry in gravity and is in obvious contrast with the claim that conformal symmetries are fake. In other words, in studying the boundary terms, we have found a nonzero conserved charge related to dilation symmetry at spatial infinity. 

Another interesting observation can be conducted in the specific case of the Schwarzschild solution with a nonzero mass $M:=\bar{\lambda}$ and $\bar{h}_{AB}=h^{(2)}_{AB} = 0 $. The dilation charge in this example is
\begin{equation}
    \begin{split}
        \mathcal{Q}^{Schw.}_{dilation}
        = & \int_{S^2} d^2x \; \sqrt{\bar{\gamma}}\frac{F}{b^2 }\left[ - 4bS- 8fW -2f\bar{\gamma}^{AB}\mathcal{L}_I \bar{\gamma}_{AB}  + M(8 b W_P+ 16F f_o + b\bar{\gamma}^{AB}\mathcal{L}_I \bar{\gamma}_{AB})\right],
    \end{split}
\end{equation}
where again the first three terms are nonvanishing contributions of the dilation symmetry and independent of the field variables. However, the remaining terms modify the ADM mass, $\mathcal{M}_{ADM} = 4\int_{S^2}d^2x \sqrt{\bar{\gamma}}f_oM$, as 
\begin{equation}
    \begin{split}
        \mathcal{M} = \int_{S^2}d^2x \sqrt{\bar{\gamma}}\left[4f_o + \frac{F}{b^2}(8 b W_P+ 16F f_o + b\bar{\gamma}^{AB}\mathcal{L}_I \bar{\gamma}_{AB})\right]M,
    \end{split}
\end{equation}
demonstrating the effect of asymptotic conformal symmetry on this physical quantity. 


\section{Discussion}\label{Conclusion and outlook}
In this paper we have included conformal symmetry at the boundary at spatial infinity. This leads to nonzero boundary terms and contributions to the conserved charges at spatial infinity. We show that the only component of the conformal symmetry that is consistent with preserving the boundary conditions is the spatial dilation. The remaining part of the spatial conformal symmetry, that is, special conformal transformations change the fall-off conditions of an asymptotically conformally flat spacetime at spatial infinity and are therefore excluded from the asymptotic symmetry generators. 

Similar to previous analyses performed at spatial infinity, we encounter contributions of divergent terms in the boundary term. In addition, divergent terms contributing from the spatial dilation symmetry add to this issue. Much the same as prior approaches to this issue, we also make use of assigning parity conditions to deal with the divergent terms. In our study of suitable parity conditions, we examine Regge-Teitelboim and Henneaux-Troessaert parity conditions in the case comprised of Poincar\'e and dilation symmetries with appropriate boundary conditions at spatial infinity. In the process of evaluating the effect of these parity conditions on our spacetime, it can be seen that the R-T parity can in fact remove all divergent terms and produce a finite nonzero charge related to dilation symmetry. However, R-T parity  also eliminates the supertranslations that we are eager to maintain. Respectively, one can argue that the significance of introducing H-T parity is to revive supertranslations. Nevertheless, the original H-T parity conditions~\cite{Henneaux-ADMBMS} are not convenient for terminating divergent terms due to conformal charges. 

In the course of introducing new parity conditions that can (i) keep the symplectic structure well defined, (ii) tailor a finite charge, and (iii) yield integrability in the charge, we prove that if one demands the parity of $\bar{\pi}^{rr}$ to be \textit{strictly} odd, any attempt to simultaneously satisfy these three conditions will recover R-T parity conditions. Nevertheless, if one uses the more satisfactory twisted H-T parity conditions~\cite{Henneaux:2018hdj}, in addition to sustaining the supertranslation charge, the divergent terms in the boundary term due to dilations also vanish and a finite dilation charge remains. This charge, expressed in (\ref{dilationcharge}), contains terms that presumably hold information about the footprints of radiation at infinity. However, this is a delicate subject and must be approached with caution; therefore we have not made any claims due to the lack of sufficient support. In the case of a Schwarzschild black hole, the ADM mass defined at spatial infinity is modified as a result of the contribution of the dilation charge. This indicates that dilation symmetry bears physical effects on the quantities defined at the boundary. As a final remark, we add that our calculation for the dilation charge also includes terms that are independent of the field variables and do not vanish, in general, hence implying that the dilations of conformal symmetry can in fact have physical effects at the boundary.  

\section*{acknowledgement}
The authors would like to thank Marc Henneaux for his insightful comments on our preprint, regarding the Henneaux-Troessaert boundary conditions, which led to further interesting results in this work.

\appendix
\begin{appendices}

\section{Asymptotic charges}\label{Appendix-Charges}
Considering the boundary term for a general variation of the smeared constraints,
\begin{equation}
    \begin{split}
        \mathcal{B}_{n}[\delta g_{ij}, \delta\pi^{ij}] = \int_{\partial\Sigma} d^2 x [-2n^i\delta\pi^r_i &+ n^r \pi^{ij}\delta g_{ij} -2\sqrt{\gamma}n\delta K \\
        & -\sqrt{\gamma}\gamma^{BC}\delta\gamma_{AC}(n K^A_B + \frac{1}{\lambda}(\partial_rn - \lambda^{C}\partial_Cn)\delta^A_B)],
    \end{split}
\end{equation}
and with the symmetry generators,
\begin{equation}
    \begin{split}
        & n = rb + f + O(r^{-1}) \\
        & n^r =r^2 Z+r F+ W+\frac{1}{r}S +  O(r^{-2}) \\
        &n^A = r J^A + Y^A + \frac{1}{r}I^A + O(r^{-2}),
    \end{split}
\end{equation}
we examine the terms individually\footnote{Although due to the boundary conditions we find $Z=0=J^A$, we include their contributions to the conserved charge in these calculations, and subsequently, we set them to zero.} in the following.

\vspace{.5cm}

\begin{center}
\textbf{Calculations for $-2n^r\delta\pi^r_r$}    
\end{center}

\begin{equation}\label{xirpirr}
    \begin{split}
        - 2 n^r \delta\pi^r_r = & - 2 n^r [\pi^{rr}\delta g_{rr} + g_{rr}\delta\pi^{rr} + \pi^{rA}\delta g_{rA} + g_{rA}\delta\pi^{rA}] \\
        = & - 2 [W(x) + r{F(x)} + r^2 Z(x) + O(\frac{1}{r})][ (\bar{\pi}^{rr} + \frac{1}{r}\pi^{(2)rr})\delta (1 + \frac{1}{r}\bar{h}_{rr} + \frac{1}{r^2}h^{(2)}_{rr}) \\
        & + (1 + \frac{1}{r}\bar{h}_{rr} + \frac{1}{r^2}h^{(2)}_{rr})\delta (\bar{\pi}^{rr} + \frac{1}{r}\pi^{(2)rr}) \\
        & + (\frac{1}{r}\bar{\pi}^{rA} + \frac{1}{r^2}\pi^{(2)rA})\delta (\bar{h}_{rA} + \frac{1}{r}h^{(2)}_{rA})\\
        & + (\bar{h}_{rA} + \frac{1}{r}h^{(2)}_{rA}) \delta (\frac{1}{r}\bar{\pi}^{rA} + \frac{1}{r^2}\pi^{(2)rA})]\\
        = & - 2[ (W + F r + r^2 Z)\delta\bar{\pi}^{rr} \\
        &  +F\delta(\bar{\pi}^{rr}\bar{h}_{rr} + \bar{\pi}^{rA}\bar{h}_{rA}+\pi^{(2)rr}) \\
        &+rZ\delta(\bar{\pi}^{rr}\bar{h}_{rr} + \bar{\pi}^{rA}\bar{h}_{rA}+\pi^{(2)rr})\\
        &+Z \delta(\bar{\pi}^{rr} h^{(2)}_{rr} + \pi^{(2)rr}\bar{h}_{rr} + \pi^{(2)rA}\bar{h}_{rA} + \bar{\pi}^{rA}h^{(2)}_{rA})],
    \end{split}
\end{equation}
where the divergent terms are $-2r^2(Z\delta\bar{\pi}^{rr})$ and $ -2r(F\delta\bar{\pi}^{rr} + Z\delta[\bar{\pi}^{rr}\bar{h}_{rr} + \bar{\pi}^{rA}\bar{h}_{rA} + \pi^{(2)rr}])$.

\vspace{.5cm}

\begin{center}
\textbf{Calculations for $- 2 n^A \delta\pi^r_A$}
\end{center}

\begin{equation}
    \begin{split}
        - 2 n^A \delta\pi^r_A = & - 2 n^A \delta[\pi^{ri}g_{iA}] \\
        = & -2 Y^A \delta[\bar{\pi}^{rB}\bar{h}_{AB} + \bar{\gamma}_{AB}\pi^{(2)rB} + \bar{h}_{rA}\bar{\pi}^{rr}] \\
        & -2 I^A[\bar{\gamma}_{AB}\delta\bar{\pi}^{rB}] \\
        & - 2 J^A \delta[\bar{\pi}^{rr} h^{(2)}_{rA} + \pi^{(2)rr}\bar{h}_{rA} + \pi^{(2)rB}\bar{h}_{AB} + \bar{\pi}^{rB}\delta h^{(2)}_{AB}] \\
        & -2 r( Y^A[\bar{\gamma}_{AB}\delta\bar{\pi}^{rB}] + J^A \delta[\bar{\pi}^{rB}\bar{h}_{AB} + \bar{\gamma}_{AB}\pi^{(2)rB} + \bar{h}_{rA}\bar{\pi}^{rr}]) \\
        & - 2 r^2 J^A [\bar{\gamma}_{AB}\delta\bar{\pi}^{rB}],
    \end{split}
\end{equation}

\vspace{.5cm}

\begin{center}
\textbf{Calculations for $+n^r\pi^{ij}\delta g_{ij}$}
\end{center}

\begin{equation}
    \begin{split}
        +n^r\pi^{ij}\delta g_{ij} = & (W + r F +  r^2 Z + O(\frac{1}{r}))\pi^{ij}\delta g_{ij} = \\
        = & (W + r F + r^2 Z + O(\frac{1}{r}))(\pi^{rr}\delta g_{rr} + \pi^{rA}\delta g_{rA} + \pi^{AB}\delta g_{AB}) \\
        = & (W + r F + r^2 Z  + O(\frac{1}{r}))([\bar{\pi}^{rr} + \frac{1}{r} \pi^{(2)rr}]\delta[1 + \frac{1}{r}\bar{h}_{rr} + \frac{1}{r^2}h^{(2)}_{rr}] \\
        & + 2[\frac{1}{r}\bar{\pi}^{rA} + \frac{1}{r^2}\pi^{(2)rA}]\delta[\bar{h}_{rA} + \frac{1}{r}h^{(2)}_{rA}]\\
        & + [\frac{1}{r^2}\bar{\pi}^{AB} + \frac{1}{r^3}\pi^{(2)AB}]\delta[r^2\bar{\gamma}_{AB} + r \bar{h}_{AB} + h^{(2)}_{AB}]) \\
        = & Z( \pi^{(2)rr}\delta\bar{h}_{rr} + \bar{\pi}^{rr} \delta h^{(2)}_{rr} + 2\pi^{(2)rA}\delta\bar{h}_{rA}  + 2\bar{\pi}^{rA}\delta h^{(2)}_{rA} + \bar{\pi}^{AB}\delta h^{(2)}_{AB} + \pi^{(2)AB}\delta\bar{h}_{AB})
        \\
        & + F(\bar{\pi}^{rr}\delta\bar{h}_{rr} + 2 \bar{\pi}^{rA}\delta\bar{h}_{rA} + \bar{\pi}^{AB}\delta\bar{h}_{AB})\\
        & + r Z (\bar{\pi}^{rr}\delta\bar{h}_{rr} + 2 \bar{\pi}^{rA}\delta\bar{h}_{rA} + \bar{\pi}^{AB}\delta\bar{h}_{AB})
    \end{split}
\end{equation}

\vspace{.5cm}

\begin{center}
\textbf{Calculations for $+ n^r \pi^{rr}\delta g_{rr}$, $+n^r \pi^{rA} \delta g_{rA}$, and $+n^r \pi^{AB}\delta g_{AB}$}
\end{center}

\begin{equation}
    \begin{split}
        + n^r \pi^{rr}\delta g_{rr} =&  W[\bar{\pi}^{rr} + \frac{1}{r} \pi^{(2)rr}]\delta[1 + \frac{1}{r}\bar{h}_{rr} + \frac{1}{r^2}h^{(2)}_{rr}] = O(\frac{1}{r}) \\
        +n^r \pi^{rA} \delta g_{rA} = & W[\frac{1}{r}\bar{\pi}^{rA} + \frac{1}{r^2}\pi^{(2)rA}]\delta[\bar{h}_{rA} + \frac{1}{r}h^{(2)}_{rA}] = O(\frac{1}{r}) \\
        +n^r \pi^{AB}\delta g_{AB} =& W[\frac{1}{r^2}\bar{\pi}^{AB} + \frac{1}{r^3}\pi^{(2)AB}]\delta[r^2\bar{\gamma}_{AB} + r \bar{h}_{AB} + h^{(2)}_{AB}] = O(\frac{1}{r})
    \end{split}
\end{equation}

\vspace{.5cm}

\begin{center}
\textbf{Calculations for $- 2 \sqrt{g}n \delta K$}
\end{center}

\begin{equation}
    \begin{split}
        - 2 \sqrt{g}n \delta K = & - 2 r^2 \sqrt{\bar{\gamma}} (rb+f)\delta(-\frac{2}{r} + \frac{1}{r^2}\bar{k} + \frac{1}{r^3}k^{(2)})) \\
        = & - 2 \sqrt{\bar{\gamma}} (rb+f)( \delta\bar{k} + \frac{1}{r}\delta k^{(2)}) )\\
        = & - 2 \sqrt{\bar{\gamma}} f \delta \bar{k} - 2 \sqrt{\bar{\gamma}} b \delta k^{(2)} - 2 \sqrt{\bar{\gamma}} r b\delta\bar{k}
    \end{split}
\end{equation}

\vspace{.5cm}

\begin{center}
\textbf{Calculations for $- \sqrt{g}g_{BC}\delta g_{AC} n K^{A}_{B} -\sqrt{g}g^{AC}\delta g_{AC} \frac{1}{\lambda}\partial_rn$ }
\end{center}

\begin{equation}\label{one}
    \begin{split}
        - \sqrt{g}g_{BC}\delta g_{AC} n K^{A}_{B} = & - r^2 \sqrt{\bar{\gamma}} (rb+f)(-\frac{1}{r}\delta^A_B + \frac{1}{r^2}\bar{k}^A_B + \frac{1}{r^3}k^{(2)A}_B) \\
        & \times(\frac{1}{r^2} \bar{\gamma}^{BC} - \frac{1}{r^3}\bar{h}^{AB})\delta(r^2\bar{\gamma}_{AC} + r \bar{h}_{AC} + h^{(2)}_{AC}) \\
        = & -  \sqrt{\bar{\gamma}} (rb+f)(-r\delta^A_B + \bar{k}^A_B + \frac{1}{r}k^{(2)A}_B) \\
        & \times(\frac{1}{r^2} \bar{\gamma}^{BC} - \frac{1}{r^3}\bar{h}^{AB})\delta(r^2\bar{\gamma}_{AC} + r \bar{h}_{AC} + h^{(2)}_{AC}) \\
        = & +  \sqrt{\bar{\gamma}} [ f(\frac{1}{r} \bar{\gamma}^{AC} -\frac{1}{r^2} \bar{h}^{AC})\delta(r \bar{h}_{AC} + h^{(2)}_{AC}) \\
        & -  f(\bar{k}^A_B)(\frac{1}{r^2} \bar{\gamma}^{BC} - \frac{1}{r^3}\bar{h}^{BC})\delta(r \bar{h}_{AC} + h^{(2)}_{AC}) \\
        & -  f(\frac{1}{r}k^{(2)A}_B)(\frac{1}{r^2} \bar{\gamma}^{BC} - \frac{1}{r^3}\bar{h}^{BC})\delta(r \bar{h}_{AC} + h^{(2)}_{AC}) \\
        & + b(\delta^A_B)( \bar{\gamma}^{BC} - \frac{1}{r}\bar{h}^{BC})\delta(r \bar{h}_{AC} + h^{(2)}_{AC}) \\
        & - b(\bar{k}^A_B)(\frac{1}{r} \bar{\gamma}^{BC} - \frac{1}{r^2} \bar{h}^{BC})\delta(r \bar{h}_{AC} + h^{(2)}_{AC}) \\
        & -  (b)(k^{(2)A}_B)(\frac{1}{r^2} \bar{\gamma}^{BC} - \frac{1}{r^3}\bar{h}^{BC})\delta(r \bar{h}_{AC} + h^{(2)}_{AC})] \\
        =& \sqrt{\bar{\gamma}} f\delta\bar{h} + \sqrt{\bar{\gamma}} b [ \delta h^{(2)} - \bar{k}^A_B \delta\bar{h}^B_A - \bar{h}^{AC}\delta \bar{h}_{AC} \\
        & + r \delta\bar{h}]\\
    \end{split}
\end{equation}

\begin{equation}\label{two}
    \begin{split}
       -\sqrt{g}g^{AC}\delta g_{AC} \frac{1}{\lambda}\partial_rn = & -r^2 \sqrt{\bar{\gamma}} b (\frac{1}{r^2}\bar{\gamma}^{AC} - \frac{1}{r^3} \bar{h}^{AC})\delta (r^2 \bar{\gamma}_{AC} + r\bar{h}_{AC} + h^{(2)}_{AC})(1 - \frac{1}{r} \bar{\lambda} + \frac{1}{r^2}\lambda^{(2)}) \\
        =& - \sqrt{\bar{\gamma}} b (\bar{\gamma}^{AC} - \frac{1}{r}\bar{h}^{AC})\delta(r\bar{h}_{AC} + h^{(2)}_{AC})(1 - \frac{1}{r} \bar{\lambda} + \frac{1}{r^2}\lambda^{(2)}) \\
        =& \sqrt{\bar{\gamma}} b[-\delta\bar{h}^{(2)} + \bar{\lambda}\delta\bar{h} +\bar{h}^{AC}\delta\bar{h}_{AC} - r\delta\bar{h}]
    \end{split}
\end{equation}
Equations (\ref{one}) and (\ref{two}) give, in total,

\begin{equation}
    \begin{split}
        - \sqrt{g}g_{BC}\delta g_{AC} n K^{A}_{B} -\sqrt{g}g^{AC}\delta g_{AC} \frac{1}{\lambda}\partial_rn = \sqrt{\bar{\gamma}} f\delta\bar{h} + \sqrt{\bar{\gamma}} b [ - \bar{k}^{AB} \delta\bar{h}_{AB} + \bar{\lambda}\delta\bar{h}].
    \end{split}
\end{equation}


\vspace{.5cm}

\begin{center}
\textbf{Calculations for $+ \sqrt{g}g^{AC}\delta g_{AC} \frac{1}{\lambda} \lambda^D \partial_D n$}
\end{center}

\begin{equation}
    \begin{split}
        + \sqrt{g}g^{AC}\delta g_{AC} \frac{1}{\lambda} \lambda^D \partial_D n = & r^2 \sqrt{\bar{\gamma}}(\frac{1}{r^2}\bar{\gamma}^{AC} - \frac{1}{r^3}\bar{h}^{AC})\delta(r^2\bar{\gamma}_{AC} + r\bar{h}_{AC} + h^{(2)}_{AC})\\
        & \times (1 - \frac{1}{r}\bar{\lambda} - \frac{1}{r^2}(\bar{\lambda}^2 -\lambda^{(2)})(\frac{1}{r^2}\bar{\lambda}^D + \frac{1}{r^3}\lambda^{(2)D})(\partial_Df + r\partial_D b) \\
        = & \sqrt{\bar{\gamma}}(\frac{1}{r^2}\bar{\gamma}^{AC} - \frac{1}{r^3}\bar{h}^{AC})\delta( r\bar{h}_{AC} + h^{(2)}_{AC})\\
        & \times (1 - \frac{1}{r}\bar{\lambda} - \frac{1}{r^2}(\bar{\lambda}^2 -\lambda^{(2)}))(\bar{\lambda}^D + \frac{1}{r}\lambda^{(2)D})(\partial_Df + r\partial_D b) \\
        = & \sqrt{\bar{\gamma}}(\bar{\lambda}^D\partial_Db) \delta\bar{h}
    \end{split}
\end{equation}

Note that we have found $Z=0=J^A$; hence to sum up, all finite nonzero terms contain
\begin{equation}
    \begin{split}
        & - 2W\delta\bar{\pi}^{rr} \\
        & + F \delta(\bar{\pi}^{rr}\bar{h}_{rr} + \bar{\pi}^{rA}\bar{h}_{rA} + \pi^{(2)rr}) \\
        & - 2Y^A \delta(\bar{\pi}^{rB} \bar{h}_{AB} + \bar{\gamma}_{AB}\pi^{(2)rB} + \bar{h}_{rA}\bar{\pi}^{rr}) \\
        & - 2 I^A(\bar{\gamma}_{AB}\delta\bar{\pi}^{rB})\\
        & - 2\sqrt{\bar{\gamma}}f\delta\bar{k} - 2\sqrt{\bar{\gamma}}b\delta k^{(2)}\\
        & + \sqrt{\bar{\gamma}}f\delta\bar{h} + \sqrt{\bar{\gamma}}(-\bar{k}^{AB}\delta\bar{h}_{AB} + \bar{\lambda}\delta\bar{h}) \\
        & + \sqrt{\bar{\gamma}}(\bar{\lambda}^C\partial_Cb)\delta\bar{h},
    \end{split}
\end{equation}
and the divergent terms are
\begin{equation}
    \begin{split}
        & + rF\delta \bar{\pi}^{rr} - 2rY^A(\bar{\gamma}_{AB}\delta\bar{\pi}^{rB}) - 2r\sqrt{\bar{\gamma}}b\delta\bar{k}.
    \end{split}
\end{equation}


\section{Preservation of the parity ponditions}\label{PC-preservation}
First, consider the following definitions:
\begin{equation}
    \begin{split}
        & {\Box^C}_{AB} = - \bar{D}^C \bar{k}_{AB} + \bar{D}_A \bar{k}^C_B + \bar{D}_B \bar{k}^C_A, \\
        & \bar{\Sigma}_{AB} = \bar{D}_J{X^J}_{AB} - \bar{D}_B{X^J}_{JC} ;\;\;\;\; \bar{\Sigma} = 2 \bar{D}^2\bar{\lambda}, \\
        & \Delta_{CAB} = -\bar{\gamma}_{AB}\bar{D}_C\bar{\lambda} + \bar{\gamma}_{CA}\bar{D}_B\bar{\lambda} + \bar{\gamma}_{CB}\bar{D}_A\bar{\lambda}, \\
        & {X^J}_{AB} = {\Box^J}_{AB} - {
        \Delta^J}_{AB} = \frac{1}{2}(\bar{\gamma}^{JI}\alpha_{IAB} - \bar{h}^{JI}\beta_{IAB}), \\
        & \Gamma^C_{AB} = -\bar{\Gamma}^C_{AB} + \frac{1}{r} {X^C}_{AB},\\
        & \beta_{CAB} = -\partial_C\bar{\gamma}_{AB} + \partial_A\bar{\gamma}_{CB} + \partial_B \bar{\gamma}_{AC}, \\
        & \alpha_{CAB} = -\partial_C\bar{h}_{AB} + \partial_A\bar{h}_{CB} + \partial_B \bar{h}_{AC}.
    \end{split}
\end{equation}

In the following we present the transformation of a few field variables needed in our study, under  hypersurface deformations.

\vspace{.5cm}

\begin{center}
\textbf{Calculations for $\delta\bar{\lambda}$}
\end{center}

\begin{equation}
    \begin{split}
        2\delta_N \bar{\lambda} = & O(\frac{1}{r})_{\delta_N g_{rr}} =   n\partial_tg_{rr} + n^r\partial_rg_{rr} + Y^C\partial_Cg_{rr} + 2g_{ri}\partial_rn^i \\
        = & \frac{2n}{\sqrt{g}}(\pi_{rr} - \frac{1}{2}\pi g_{rr}) + n^r\partial_r g_{rr} + Y^C\partial_C g_{rr} \\
        = & \frac{2(rb+f)}{r^2\sqrt{\bar{\gamma}}}(g_{ri}g_{rj}\pi^{ij} - \frac{1}{2}g_{ij}\pi^{ij} g_{rr}) + n^r\partial_r g_{rr} + Y^C\partial_C g_{rr} \\
        = & \frac{2(rb+f)}{r^2\sqrt{\bar{\gamma}}}(g_{rr}g_{rr}\pi^{rr} + g_{rr}g_{rC}\pi^{rC} + g_{rC}g_{rE}\pi^{CE} \\
        & - \frac{1}{2}(g_{rr}\pi^{rr}+2g_{rC}\pi^{rC}+g_{CE}\pi^{CE}) g_{rr}) + n^r\partial_r g_{rr} + Y^C\partial_C g_{rr} \\
        = & \frac{1}{r}[ \frac{2b}{\sqrt{\bar{\gamma}}}(\frac{1}{2}\bar{\pi}^{rr} - \frac{1}{2}\bar{\pi})] + \frac{1}{r}(Y^C\partial_C\bar{h}_{rr}) \\
    \end{split}
\end{equation}
which gives
\begin{equation}\label{HD-of-Lam}
    \begin{split}
        \delta_N \bar{h}_{rr} = \frac{b}{\sqrt{\bar{\gamma}}}(\bar{\pi}^{rr} - \bar{\pi}^A_A) + Y^C\partial_C\bar{h}_{rr}.
    \end{split}
\end{equation}

\vspace{.5cm}

\begin{center}
\textbf{Calculations for $\delta\bar{\pi}^{rr}$}
\end{center}

Considering $\delta\bar{\pi}^{rr} = O(1)|_{\delta_N \pi^{rr}}$, we have
\begin{equation}
    \begin{split}
        \delta_N\pi^{rr} = & - n r^2 \sqrt{\bar{\gamma}} ({^{(3)}R}^{rr} - \frac{1}{2} g^{rr}\; {^{(3)}R}) + \frac{1}{2}\frac{n}{r^2\sqrt{\bar{\gamma}}} (\pi_{ij}\pi^{ij} - \frac{1}{2}\pi^2)g^{rr} \\
        & - \frac{2n}{r^2\sqrt{\bar{\gamma}}}(\pi^{ri}{\pi_i}^r - \frac{1}{2}\pi^{rr}\pi) + r^2\sqrt{\bar{\gamma}}(D^rD^rn - g^{rr}D_iD^in) \\
        & + \mathcal{L}_{\textbf{n}}\pi^{rr} \\
        & = - (rb+f)r^2\sqrt{\bar{\gamma}}[{^{(3)}G^{rr}}] + r^2\sqrt{\bar{\gamma}}[D^rD^r(rb+f) - \frac{1}{2}g^{rr}D_iD^i(rb+f)] + \left(\mathcal{L}_{Y}\bar{\pi}^{rr} - F\bar{\pi}^{rr}\right) + \dots
    \end{split}
\end{equation}
where the contributing terms are
\begin{equation}
    \begin{split}
        {^{(3)}G_{rA}} =& {^{(3)}R_{rA} } - \frac{1}{2}g_{rA}\; {^{(3)}R } = O(\frac{1}{r^2})
    \end{split}
\end{equation}
\begin{equation}
    \begin{split}
        {^{(3)}G_{rr}} =& {^{(3)}R_{rr} } - \frac{1}{2}g_{rr}\; {^{(3)}R } \\
        = & + (1 + \frac{\bar{\lambda}}{r} + \dots)(\frac{2}{r^2} - \frac{2\bar{k}}{r^3} + \dots) + \dots - (1+\frac{\bar{\lambda}}{r} + \dots)^2(-\frac{\delta^A_B}{r} + \frac{\bar{k}^A_B}{r^2} + \dots)(-\frac{\delta^B_A}{r}+ \frac{\bar{k}^B_A}{r^2}+..) \\
        & - (1 + \frac{\bar{\lambda}}{r} + \dots)(\frac{\bar{\gamma}^{AB}}{r^2}-\frac{\bar{h}^{AB}}{r^3}+ \dots)(\frac{\bar{D}_A\bar{D}_B\bar{\lambda}}{r} +\frac{{X^C}_{AB}\bar{D}_C\bar{\lambda}}{r^2} + \dots)\\
        & - \frac{1}{2}(1 + \frac{2\bar{\lambda}}{r}+\dots)\left[ 2(1-\frac{\bar{\lambda}}{r}+\dots)(\frac{2}{r^2} - \frac{2\bar{k}}{r^3}+\dots+N.C+\dots) + \frac{\bar{R}}{r^2} + \frac{\bar{\Sigma}}{r^3} - \frac{\bar{h}^{AB}\bar{R}_{AB}}{r^3} \right. \\
        & \left. - (-\frac{\delta^A_B}{r} + \frac{\bar{k}^A_B}{r^2}+\dots)(-\frac{\delta^B_A}{r} + \frac{\bar{k}^B_A}{r^2}+\dots) - ( -\frac{2}{r} + \frac{\bar{k}}{r^2}+\dots)^2 - 2 (1-\frac{\bar{\lambda}}{r} + \dots)(\frac{\bar{D}^2\bar{\lambda}}{r^3}+ \dots)\right]
        \\
        = & \frac{1}{r^3}[ -2\bar{\lambda} - \frac{\bar{\Sigma}}{2} ] + \dots
    \end{split}
\end{equation}

\begin{equation}
    \begin{split}
        {^{(3)}G^{rr}} = \frac{1}{r^3}\left[ -2\bar{\lambda} - \frac{\bar{\Sigma}}{2} \right] + \dots = \frac{1}{r^3}[-2\bar{\lambda} - \bar{D}^2\bar{\lambda}] + \dots
    \end{split}
\end{equation}
 
\begin{equation}
    \begin{split}
        r^2\sqrt{\bar{\gamma}} [ D^rD^r (rb + f) - g^{rr} D_iD^i(rb+f)] =& r^2\sqrt{\bar{\gamma}}[(g^{ri}g^{rj} - g^{rr}g^{ij}) D_iD_j(rb+f)] \\
        =& r^2 \sqrt{\bar{\gamma}} [ (g^{rr}g^{rr} - g^{rr}g^{rr})D_rD_r(rb+f) \\
        & + 2(g^{rr}g^{rC} - g^{rr}g^{rC}) D_rD_C(rb+f) \\
        & + (g^{rC}g^{rE} - g^{rr}g^{CE})D_CD_E(rb+f)]\\
        = & - \sqrt{\bar{\gamma}}(\bar{D}^2f - {X^{JC}}_C\partial_Jb - 6 b\bar{\lambda} + b\bar{k}) \\
        & \sqrt{\bar{\gamma}}(-\bar{D}^2f + 2\bar{D}_E\bar{k}^{CE}\partial_Cb - \bar{D}^C\bar{k}\partial_Cb - 6b\bar{\lambda} + b\bar{k})
    \end{split}
\end{equation}
which finally gives
\begin{equation}
    \begin{split}
        \delta\bar{\pi}^{rr} = & \sqrt{\bar{\gamma}}\left[-b(-2\bar{\lambda} - \bar{D}^2\bar{\lambda}) -\bar{D}^2f + 2(\bar{D}_E\bar{k}^{EC})(\bar{D}_Cb) - (\bar{D}^C\bar{k})(\bar{D}_Cb) + 6b\bar{\lambda} - b \bar{k}\right] + \mathcal{L}_{Y}\bar{\pi}^{rr} - F\bar{\pi}^{rr}.
    \end{split}
\end{equation}

\vspace{.5cm}

\begin{center}
\textbf{Calculations for $\delta\bar{\pi}^{rA}$}
\end{center}

Noting $\delta\bar{\pi}^{rA} = O(r^{-1})|_{\delta_N\pi^{rA}}$, we have
\begin{equation}
    \begin{split}
        \delta_N\pi^{rA} = & - n r^2 \sqrt{\bar{\gamma}} ({^{(3)}R}^{rA} - \frac{1}{2} g^{rA}\; {^{(3)}R}) + \frac{1}{2}\frac{n}{r^2\sqrt{\bar{\gamma}}} (\pi_{ij}\pi^{ij} - \frac{1}{2}\pi^2)g^{rA} \\
        & - \frac{2n}{r^2\sqrt{\bar{\gamma}}}(\pi^{ri}{\pi_i}^A - \frac{1}{2}\pi^{rA}\pi) + r^2\sqrt{\bar{\gamma}}(D^rD^An - g^{rA}D_iD^in) 
        + \mathcal{L}_{\textbf{n}}\pi^{rA } \\
        & = - (rb+f)r^2\sqrt{\bar{\gamma}}[{^{(3)}G^{rA}}] + r^2\sqrt{\bar{\gamma}}[D^rD^A(rb+f) - \frac{1}{2}g^{rA}D_iD^i(rb+f)] + \frac{1}{r}\left(\mathcal{L}_{Y}\bar{\pi}^{rA} - F\bar{\pi}^{rA}\right) + \dots
    \end{split}
\end{equation}

The contributing terms are

\begin{equation}
    \begin{split}
        {^{(3)}G}^{rA} = & {^{(3)}R}^{rA} - \frac{1}{2}g^{rA}\;{^{(3)}R} = g^{ri}g^{Aj}\;{^{(3)}R}_{ij} - \frac{1}{2}g^{rA}\;{^{(3)}R} \\
        = & g^{rr}g^{Ar}{^{(3)}R}_{rr} + (g^{rr}g^{AB}+g^{rB}g^{Ar}){^{(3)}R}_{rB} + (g^{rB}g^{AC}+g^{rC}g^{AB}){^{(3)}R}_{BC}
        \\ 
        & = g^{rr}g^{AB}\;{^{(3)}R}_{rB} + O(r^{-5}) \\
        = & (1 - \frac{1}{\bar{\lambda}} + \dots)(\frac{\bar{\gamma}^{AB}}{r^2} - \frac{\bar{h}^{AB}}{r^3} + \dots)(\partial_BK - D_CK^C_B) + O(\frac{1}{r^5}) =  \frac{1}{r^4}(\bar{D}^A\bar{k}  - \bar{D}_C\bar{k}^{CA}) + O(r^{-5})
    \end{split}
\end{equation}
and 
\begin{equation}
    \begin{split}
        r^2\sqrt{\bar{\gamma}}(D^rD^An - & g^{rA}D_iD^in) =  r^2\sqrt{\bar{\gamma}}g^{rr}g^{AB}D_rD_Bn + O(r^{-2})\\
        = &r^2\sqrt{\bar{\gamma}}\left(1-\frac{2\bar{\lambda}}{r}+\dots \right)\left( \frac{\bar{\gamma}^{AB}}{r^2} + \dots \right)\left( \partial_r\partial_B(rb+f) - \Gamma^r_{rB}\partial_r(rb+f) - \Gamma^C_{rB}\partial_C(rb+f)\right) \\
        = & \sqrt{\bar{\gamma}}(\bar{\gamma}^{AB} + \dots)(\partial_Bb - b(1-\frac{\bar{\lambda}}{r}+\dots)(\frac{\partial_B\bar{\lambda}}{r} + \dots) + \partial_C(rb+f)(1+\frac{\bar{\lambda}}{r}+\dots)(-\frac{\delta^C_A}{r}+\frac{\bar{k}^C_A}{r^2}+\dots))\\
        = & \sqrt{\bar{\gamma}}(\bar{D}^Ab + \frac{1}{r}(-b\bar{D}^A\bar{\lambda} - \bar{D}^Af + \bar{k}^{CA}\bar{D}_Cb - \bar{\lambda}\bar{D}^Ab)) + O(r^{-2})
    \end{split}
\end{equation}
Therefore, 
\begin{equation}\label{HD-of-pi-rA}
    \begin{split}
        \delta\bar{\pi}^{rA} = &  b\sqrt{\bar{\gamma}}(\bar{D}_C\bar{k}^{CA} - \bar{D}^A\bar{k}-b\bar{D}^A\bar{\lambda}) + \sqrt{\bar{\gamma}}( - \bar{D}^Af + \bar{k}^{CA}\bar{D}_Cb - \bar{\lambda}\bar{D}^Ab)) \\
        = & \sqrt{\bar{\gamma}}\left[\bar{D}_C(b\bar{k}^{CA} - \bar{D}^A(b\bar{\lambda}) - b\bar{D}^A\bar{k} - \bar{D}^Af\right] + \mathcal{L}_Y\bar{\pi}^{rA} - F\bar{\pi}^{rA}.
    \end{split}
\end{equation}

\vspace{.5cm}

\begin{center}
\textbf{Calculations for $\delta\bar{\pi}^{AB}$}
\end{center}

For the next field variable $\delta\bar{\pi}^{AB} = O(r^{-2})_{\delta\pi^{AB}}$, the calculations are
\begin{equation}
    \begin{split}
        \delta_N\pi^{AB} = & - (rb+f) r^2 \sqrt{\bar{\gamma}} (\;{^{(3)}R}^{AB} - \frac{1}{2} g^{AB} {^{(3)}R}) + \frac{1}{2}\frac{rb+f}{r^2\sqrt{\bar{\gamma}}}(\pi_{mn}\pi^{mn} - \frac{1}{2}\pi^2)g^{AB} \\
        & - \frac{2(rb+f)}{r^2\sqrt{\bar{\gamma}}} (\pi^{Am}{\pi_m}^B - \frac{1}{2}\pi \pi^{AB}) + r^2 \sqrt{\bar{\gamma}} ( D^AD^B (rb+f) - g^{AB} D_mD^m(rb+f))\\
        & + \mathcal{L}_{\textbf{n}}\pi^{AB} \\
    \end{split}
\end{equation}

The contributing terms are
\begin{equation}
    \begin{split}
        {^{(3)}G^{AB}} = & g^{Ai}g^{Bj}G_{ij} = g^{Ar}g^{Br}G_{rr} + 2g^{Ar}g^{BC}G_{rC} + g^{AC}g^{BE} {^{(3)}G_{CE}} \\
        = & (\frac{\bar{\gamma}^{AC}}{r^2} - \frac{\bar{h}^{AC}}{r^3}+\dots)(\frac{\bar{\gamma}^{BE}}{r^2} - \frac{\bar{h}^{BE}}{r^3}+\dots)\left[\frac{1}{r}\{-\bar{D}_C\bar{D}_E\bar{\lambda}+\bar{\gamma}_{CE}\bar{D}^2\bar{\lambda}\} + {^{\gamma}R_{CE}} - \frac{1}{2}g_{CE}{^{\gamma}R} \right]\\
        = & {^\gamma G^{AB}} + \frac{1}{r^5}(\bar{\gamma}^{AB}\bar{D}^2\bar{\lambda} - \bar{D}^A\bar{D}^B\bar{\lambda}) + \dots,
    \end{split}
\end{equation}

\begin{equation}
    \begin{split}
        ^{(3)}R_{AB} = &  \frac{1}{\lambda}\partial_r K_{AB} + 2 K_{AC}K^C_B - KK_{AB} - \frac{1}{\lambda}D_AD_B\lambda \\
        & + {^{\gamma}R_{AB}} - \frac{1}{\lambda}\mathcal{L}_{\lambda}K_{AB} \\
        = & (1 - \frac{\bar{\lambda}}{r} + \dots) ( - \bar{\gamma}_{AB}+\dots) + 2(-r\bar{\gamma}_{AC} - \bar{h}_{AC} + \bar{k}_{AC} + \dots)(-\frac{\delta^C_B}{r} + \frac{\bar{k}^C_B}{r^2}+\dots) \\
        & - (-\frac{2}{r} + \frac{\bar{k}}{r^2} + \dots) (-r\bar{\gamma}_{AB} - \bar{h}_{AB} + \bar{k}_{AB} + \dots) - (1 - \frac{\lambda}{r} + \dots)(\frac{\bar{D}_A\bar{D}_B\bar{\lambda}}{r}+ \dots) \\
        & + {^{\gamma}R_{AB}} - (1-\frac{\bar{\lambda}}{r}+\dots) (\lambda^C\partial_C K_{AB} + K_{AC}\partial_B\lambda^C + K_{BC}\partial_A\lambda^C) \\
        = & - \bar{\gamma}_{AB} + \frac{1}{r}\left[\bar{\lambda}\bar{\gamma}_{AB} - 2 \bar{k}_{AB} + \bar{k}\bar{\gamma}_{AB} - \bar{D}_A\bar{D}_B\bar{\lambda}\right] + {^{\gamma}R_{AB}} + \dots, \\
        -\frac{1}{2}g_{AB}{^{(3)}R} = & - \frac{1}{2}(r^2 \bar{\gamma}_{AB} + r\bar{h}_{AB}+\dots) \left[ 2(1-\frac{\bar{\lambda}}{r} + \dots)(\frac{2}{r^2} - \frac{2\bar{k}}{r^3}+\dots) + {^{\gamma}R} \right. \\
        & \left. - (-\frac{\delta^A_B}{r} + \frac{\bar{k}^A_B}{r^2}+\dots)(-\frac{\delta^B_A}{r} + \frac{\bar{k}^B_A}{r^2}+\dots) - (-\frac{2}{r}+\frac{\bar{k}}{r^2}+\dots)^2 - (2(1-\frac{\bar{\lambda}}{r}+\dots)(\frac{\bar{D}_C\bar{D}^C\bar{\lambda}}{r^3}+\dots) \right] \\
        = & \frac{2\bar{\lambda}}{r}\bar{\gamma}_{AB} + \bar{\gamma}_{AB} - \frac{\bar{k}\bar{\gamma}_{AB}}{r} - \frac{\bar{h}_{AB}}{r} + \frac{\bar{\gamma}_{AB}}{r}\bar{D}_C\bar{D}^C\bar{\lambda} - \frac{1}{2}g_{AB}{^{\gamma}R} \\
        = & \bar{\gamma}_{AB} + \frac{1}{r}\left[2 \bar{k}_{AB} - \bar{k}\bar{\gamma}_{AB} - \bar{\lambda}\bar{\gamma}_{AB} + \bar{\lambda}_{AB}\bar{D}^2\bar{\lambda}\right] - \frac{1}{2}g_{AB}{^{\gamma}R} + \dots,
    \end{split}
\end{equation}

\begin{equation}
    \begin{split}
        {^{\gamma}R_{AB}} = & \partial_C{^{\gamma}\Gamma^C_{AB}} - \partial_B {^{\gamma}\Gamma^C_{CA}} + {^{\gamma}\Gamma^C_{AB}}{^{\gamma}\Gamma^E_{EC}} - {^{\gamma}\Gamma^C_{AE}}{^{\gamma}\Gamma^E_{BC}} \\
        = & \bar{R}_{AB} + \frac{1}{2r}\left[2\partial_C( {\Box^C}_{AB} - {\Delta^C}_{AB} ) - 2 \partial_B( {\Box^C}_{CA} - {\Delta^C}_{CA} ) \right. \\
        &  + 2( {\Box^C}_{AB} - {\Delta^C}_{AB} )\bar{\Gamma}^E_{EC} + 2 ( {\Box^E}_{EC} - {\Delta^E}_{EC} )\bar{\Gamma}^C_{AB}\\
        & \left. - 2 ( {\Box^C}_{AE} - {\Delta^C}_{AE} ) \bar{\Gamma}^E_{BC} - 2( {\Box^E}_{BC} - {\Delta^E}_{BC} ) \bar{\Gamma}^C_{AE} \right] \\
        = & \bar{R}_{AB} + \frac{1}{r}\left[\bar{D}_C( {\Box^C}_{AB} - {\Delta^C}_{AB} ) - \bar{D}_B( {\Box^C}_{CA} - {\Delta^C}_{CA} )\right]+\dots, \\
        -\frac{1}{2}g_{AB} g^{CE}\;{^\gamma R_{CE}} = & - \frac{1}{2}(r^2 \bar{\gamma}_{AB} + r\bar{h}_{AB} + \dots)(\frac{\gamma^{CE}}{r^2} - \frac{\bar{h}^{CE}}{r^3} + \dots)\;{^\gamma R_{CE}} \\
        = & -\frac{1}{2}\bar{\gamma}_{AB}\bar{R} - \frac{1}{2r}\left[(\bar{h}_{AB}\bar{\gamma}^{CE} - \bar{h}^{CE}\bar{\gamma}_{AB}) \bar{R}_{CE} + \bar{\gamma}_{AB}\bar{\Sigma}\right]\\
        = & -\frac{1}{2}\bar{\gamma}_{AB}\bar{R} - \frac{1}{2r}\left[2\bar{k}_{AB}\bar{R} - 2\bar{k}^{CE}\bar{\gamma}_{AB} \bar{R}_{CE} + \bar{\gamma}_{AB}\bar{\Sigma}\right]+\dots.
    \end{split}
\end{equation}

Using $\bar{R}_{AB}=\bar{\gamma}_{AB}$ and $\bar{R} = 2$ we have
\begin{equation}
    \begin{split}
        G^{AB} = & g^{AJ}g^{BI}G_{IJ} \\
        = & (\frac{\bar{\gamma}^{AJ}}{r^2} - \frac{\bar{h}^{AJ}}{r^3}+\dots)(\frac{\bar{\gamma}^{BI}}{r^2} - \frac{\bar{h}^{BI}}{r^3}+\dots)(R_{IJ} - \frac{1}{2}g_{IJ}R) \\
        = & ( \bar{R}^{AB} - \frac{1}{2}\bar{\gamma}^{AB}\bar{R}) \\
        & + \frac{1}{r^5}\left[ (-\bar{\gamma}^{AJ}\bar{h}^{BI} - \bar{h}^{AJ}\bar{\gamma}^{BI})(\bar{R}_{IJ} - \frac{1}{2}\bar{\gamma}_{IJ}\bar{R}) \right. \\
        & \left. + \bar{\gamma}^{AJ}\bar{\gamma}^{BI}(\bar{\Sigma}_{JI} - \bar{R}\bar{k}_{JI} + \bar{\gamma}_{JI}\bar{k}^{CE}\bar{R}_{CE} - \frac{1}{2}\bar{\Sigma}\bar{\gamma}_{JI})\right] \\
        = & \frac{1}{r^5}\left[\bar{\Sigma}^{AB} - \frac{1}{2}\bar{\Sigma} - 2\bar{k}^{AB} - \bar{\gamma}^{AB}\bar{k} \right] + O(r^{-6})
    \end{split}
\end{equation}
Also, 
\begin{equation}
    \begin{split}
        &r^2\sqrt{\bar{\gamma}} [D^AD^Bn - g^{AB}D_iD^in] = r^2\sqrt{\bar{\gamma}} (D_iD_jn)[g^{Ai}g^{Bj} - g^{AB}g^{ij}] \\
        = &  \frac{\sqrt{\bar{\gamma}}}{r^2}[\bar{\lambda}\bar{\gamma}^{AB} - \bar{\gamma}^{AB}\bar{D}_C\bar{\lambda}\bar{D}^Cb + \bar{D}^A\bar{D}^Bf - \bar{\gamma}^{AB}\bar{D}^2f \\
        & - (\Box^{JAB} - \bar{\gamma}^{AB}{\Box^{JC}}_C - \Delta^{JAB} + \bar{\gamma}^{AB}{\Delta^{JC}}_C)\partial_Jb + b\bar{k}^{AB} - b\bar{k}\bar{\gamma}^{AB} + 3b\bar{\lambda}\bar{\gamma}^{AB}]\\
        = & \frac{\sqrt{\bar{\gamma}}}{r^2}[\bar{D}^A\bar{D}_Bf - \bar{\gamma}^{AB}\bar{D}^2f + b\bar{\lambda}\bar{\gamma}^{AB} - \bar{\gamma}^{AB}\bar{D}_Cb\bar{D}^C\bar{\lambda} \\
        & + (\bar{D}^J\bar{k}^{AB} - \bar{D}^A\bar{k}{JB} - \bar{D}^B\bar{k}^{JA} - \bar{\gamma}^{AB}\bar{D}^J\bar{k} + 2\bar{\gamma}^{AB}\bar{D}_C\bar{k}^{CJ}\\
        & + \bar{\gamma}^{AB}\bar{D}^J\bar{\lambda} - \bar{\gamma}^{AJ}\bar{D}^B\bar{\lambda} - \bar{\gamma}^{BJ}\bar{D}^A\bar{\lambda})\partial_Jb]
    \end{split}
\end{equation}

Finally,  
\begin{equation}
    \begin{split}
        \delta\bar{\pi}^{AB} & = \sqrt{\bar{\gamma}}\left[ -\bar{\gamma}^{AB}\bar{D}^2\bar{\lambda} + b \bar{D}^A\bar{D}^B \bar{\lambda} - b \bar{D}_J \Box^{JAB} + b \bar{D}^A{\Box_J}^{JB} + 2 b \bar{k}^{AB} - b \bar{k}\bar{\gamma}^{AB} \right.\\
        & + \frac{1}{2} b \bar{\gamma}^{AB}(\bar{D}_J{\Box^{JC}}_C -\bar{D}_J {\Box_C}^{CJ}) + \bar{\gamma}^{AB}{\Box^{JC}}_C(\partial_J b) - \Box^{JAB}(\partial_J b) \\
        & + (\bar{D}^A\bar{D}^B f - \bar{\gamma}^{AB}\bar{D}^2 f) + b\bar{\lambda}\bar{\gamma}^{AB} - \bar{\gamma}^{AB}(\bar{D}_C b)(\bar{D}^C\bar{\lambda}) \\
        &\left. - \bar{\gamma}^{AB}(\bar{D}^Cb)(\bar{D}^C\bar{\lambda}) + (\bar{D}^A\bar{\lambda})(\bar{D}^Bb) + (\bar{D}^Ab )(\bar{D}^B\bar{\lambda}) \right] + \mathcal{L}_Y\bar{\pi}^{AB}.
    \end{split}
\end{equation}

\section{Calculations for dilation charge}\label{pi^rr}

We have
\begin{equation}
    \begin{split}
        \pi^{rr} = & \sqrt{g}(K^{rr} - g^{rr}K) = r^2 \frac{\sqrt{\bar{\gamma}}}{2n} (g^{ri}g^{rj}-g^{rr}g^{ij})(\dot{g}_{ij} - \mathcal{L}_{\textbf{n}}g_{ij})\\
        = & \sqrt{\bar{\gamma}} (\frac{r}{2b}- \frac{f}{b^2} + \dots) (- \frac{g^{AB}}{\lambda^2}+\dots)(\dot{g}_{AB} - \mathcal{L}_{\textbf{n}}g_{AB})\\
        = & -\sqrt{\bar{\gamma}} (\frac{\bar{\gamma}^{AB}}{2br} - \frac{\bar{k}^{AB}}{br^2} - \frac{h^{(2)AB} + 2\bar{\lambda}\bar{h}^{AB} - 3\bar{\lambda}^2\bar{\gamma}^{AB} }{2br^3} + \frac{f\bar{\gamma}^{AB}}{b^2r^2} - \frac{2f\bar{k}^{AB}}{b^2r^3} + \dots )\times \\
        &\times(-2r^2 F\bar{\gamma}_{AB} - r( \dot{\bar{h}}_{AB} + 2 W \bar{\gamma}_{AB} + F\bar{h}_{AB}+ \mathcal{L}_Y\bar{h}_{AB} + \mathcal{L}_I\bar{\gamma}_{AB}) \\
        & \;\; + \dot{h}^{(2)}_{AB} - 2S\bar{\gamma}_{AB} - W\bar{h}_{AB} - \mathcal{L}_Y  h^{(2)}_{AB} - \mathcal{L}_I\bar{h}_{AB}+ \dots)\\
        = & \frac{\sqrt{\bar{\gamma}}}{2b}\left[\bar{\gamma}^{AB}\dot{\bar{h}}_{AB} + 4 W + F\bar{h}+ {\gamma}^{AB}\mathcal{L}_Y\bar{h}_{AB} + {\gamma}^{AB}\mathcal{L}_I\bar{\gamma}_{AB}\right] \\
        & -\frac{\sqrt{\bar{\gamma}}}{2b^2 r}\left[ b(\bar{\gamma}^{AB}\dot{h}^{(2)}_{AB} - 4S - W\bar{h} - {\gamma}^{AB}\mathcal{L}_Y  h^{(2)}_{AB} - {\gamma}^{AB}\mathcal{L}_I\bar{h}_{AB}) \right.\\
        & - (2f\bar{\gamma}^{AB} - 2b\bar{k}^{AB})( \dot{\bar{h}}_{AB} + 2 W \bar{\gamma}_{AB} + F\bar{h}_{AB}+ \mathcal{L}_Y\bar{h}_{AB} + \mathcal{L}_I\bar{\gamma}_{AB}) \\
        & \left.+ 2F(b\bar{\gamma}_{AB}h^{(2)AB} + 2b\bar{\lambda}\bar{h} - 6b\bar{\lambda}^2 + 4f\bar{k})\right] + O(r^{-2})
    \end{split}
\end{equation}
Therefore,
\begin{equation}
    \begin{split}
        \bar{\pi}^{(2)rr}=&-\frac{\sqrt{\bar{\gamma}}}{2b^2}\left[ b(\bar{\gamma}^{AB}\dot{h}^{(2)}_{AB} - 4S - W\bar{h} - {\gamma}^{AB}\mathcal{L}_Y  h^{(2)}_{AB} - {\gamma}^{AB}\mathcal{L}_I\bar{h}_{AB}) \right.\\
        & - (2f\bar{\gamma}^{AB} - 2b\bar{k}^{AB})( \dot{\bar{h}}_{AB} + 2 W \bar{\gamma}_{AB} + F\bar{h}_{AB}+ \mathcal{L}_Y\bar{h}_{AB} + \mathcal{L}_I\bar{\gamma}_{AB}) \\
        & \left.+ 2F(b\bar{\gamma}_{AB}h^{(2)AB} + 2b\bar{\lambda}\bar{h} - 6b\bar{\lambda}^2 + 4f\bar{k})\right].
    \end{split}
\end{equation}

\end{appendices}

\bibliography{ref}

\end{document}